\documentclass[nofootinbib,amsmath,amssymb,10pt,eqsecnum, twocolumn,a4paper]{revtex4-1}

\usepackage{graphicx}
\usepackage{rotating}
\usepackage{amsmath}
\usepackage{amsfonts}
\usepackage{amssymb}
\usepackage{float}
\usepackage{color}
\usepackage[breaklinks, colorlinks, citecolor=blue]{hyperref}
\usepackage{natbib,hyperref,ifthen}
\newcommand{\dint} {\displaystyle\int}
\newcommand{\dsum} {\displaystyle\sum}

\usepackage{array}
\newcolumntype{R}[1]{>{\raggedleft\let\newline\\\arraybackslash\hspace{0pt}}m{#1}}
\newcolumntype{L}[1]{>{\raggedright\let\newline\\\arraybackslash\hspace{0pt}}m{#1}}
\usepackage[vmargin=2cm,hmargin=2cm]{geometry}
\begin{document}

    \title{CMB constraints on cosmic strings and superstrings}
    \author{Tom Charnock}
    \email{tom.charnock@nottingham.ac.uk}
    \affiliation{Centre for Astronomy and Particle Theory, University of Nottingham, University Park, Nottingham, NG7 2RD, U.K.}
    \author{Anastasios Avgoustidis}
    \email{anastasios.avgoustidis@nottingham.ac.uk}
    \affiliation{Centre for  Astronomy and Particle Theory, University of Nottingham, University Park, Nottingham, NG7 2RD, U.K.}
    \author{Edmund J. Copeland}
    \email{ed.copeland@nottingham.ac.uk}
    \affiliation{Centre for  Astronomy and Particle Theory, University of Nottingham, University Park, Nottingham, NG7 2RD, U.K.}
    \author{Adam Moss}
    \email{adam.moss@nottingham.ac.uk}
    \affiliation{Centre for  Astronomy and Particle Theory, University of Nottingham, University Park, Nottingham, NG7 2RD, U.K.}

    \begin{abstract}
                We present the first complete Markov chain Monte Carlo analysis of cosmological models with evolving cosmic \mbox{(super)}string networks, using the unconnected segment model in the unequal-time correlator formalism. For ordinary cosmic string networks, we derive joint constraints on $\Lambda$ cold dark matter (CDM) and string network parameters, namely the string tension $G\mu$, the loop-chopping efficiency $c_{\rm r}$, and the string wiggliness $\alpha$. For cosmic superstrings, we obtain joint constraints on the fundamental string tension $G\mu_{\rm F}$, the string coupling $g_{\rm s}$, the self-interaction coefficient $c_{\rm s}$, and the volume of compact extra dimensions $w$. This constitutes the most comprehensive CMB analysis of $\Lambda$CDM cosmology + strings to date. For ordinary cosmic string networks our updated constraint on the string tension, obtained using Planck2015 temperature and polarisation data, is $G\mu<1.1\times10^{-7}$ in relativistic units, while for cosmic superstrings our constraint on the fundamental string tension after marginalising over $g_{\rm s}$, $c_{\rm s}$, and $w$ is $G\mu_{\rm F}<2.8\times10^{-8}$. 
    \end{abstract}  
      
    \maketitle
    
 \section{Introduction}
    
        \indent Cosmic strings are linelike concentrations of energy that can arise as topological defects in theories of the early Universe~\cite{Kibble80,Vilenkin:2000,Hindmarsh93,Copeland:2009ga,Copeland:2011dx}. In particular, they form naturally in models of hybrid inflation~\cite{Baumann:2007ah,Burgess:2001fx,Copeland:1994vg,Dvali:1994ms,Dvali:1998pa,Kachru:2003sx,Linde:1993cn} in which the inflationary phase ends with a second-order phase transition~\cite{Burgess:2001fx,Copeland:2003bj,Jeannerot:2003qv,Sarangi:2002yt}. Although they were originally considered as an alternative candidate for providing the seeds for structure formation in the Universe~\cite{Brandenberger:1990,Contaldi:1998mx,Kibble:1985,Pen:1997ae}, it is now understood that they cannot give rise to the observed acoustic peak structure in the  power spectrum~\cite{Albrecht:1997nt,Albrecht:1997mz,Avelino:1997hy,Battye:1997hu,Copeland:1999gn}, but can play a subdominant role. There is a wide range of potential observational signatures of cosmic strings, for example, linelike discontinuities in the cosmic microwave background (CMB) temperature anisotropy via the Kaiser-Stebbins effect~\cite{Gott:1984ef,Kaiser:1984iv}. Thus, strings provide a powerful tool for testing theories of the early Universe. Observations have strongly constrained the contribution of cosmic strings to the total CMB anisotropy~\cite{Albrecht:1997nt,Pogosian:2003mz,Wyman:2005tu,Battye:2010xz,Dvorkin:2011aj,Shlaer:2012,Ade:2013xla}. Current data place a 2$\sigma$ upper bound on the string tension of $G\mu<1.3\times10^{-7}$ for Nambu-Goto strings~\cite{Ade:2015xua} or $G\mu<2.7\times10^{-7}$ for Abelian-Higgs strings~\cite{Lizarraga:2014xza}, which corresponds to $\sim$1\% of the total temperature anisotropy at $\ell=10$. $G$ is the gravitational constant, $\mu$ is the tension of the string and $c=1$ in relativistic units. Although this may seem insignificant, there is still constraining power left in the data since strings generate specific signatures in the primordial B-mode polarisation spectrum~\cite{Bevis:2007qz,Mukherjee:2010ve,Pogosian:2003mz,Pogosian:2007gi,Seljak:1997ii,Seljak:2006hi,Urrestilla:2008jv}, which can now be analysed with the Planck2015 polarisation~\cite{Aghanim:2015xee} and joint BICEP2 data~\cite{Ade:2015tva}.
        \\*
        \\*
        \indent Going beyond the simplest cosmic string models, complex networks of multiple types of interacting superstrings, each with a different tension, can also be considered. Notably, interacting networks of fundamental F-strings, one-dimensional D-branes (D-strings), and bound (FD) states between F- and D-strings, collectively referred to as cosmic superstrings, arise naturally in string theoretic inflation~\cite{Burgess:2001fx,Dvali:2003,Tye:2005fn}. These networks are notably different to their simpler, single-type string counterparts since the different string types have intercommutation probabilities that are not necessarily unity~\cite{Jones:2003da,Jackson:2004zg,Hanany:2005bc,Jackson:2007hn,Avgoustidis:2005nv,Tye:2005fn,Avgoustidis:2007aa}. The interactions among different string types are also much more complex, as colliding strings can zip together or unzip, producing heavier or lighter FD-string states carrying different charges. These features affect CMB signatures allowing us to obtain constraints on string theory parameters such as the string coupling $g_{\rm s}$ and the fundamental string tension $\mu_{\rm F}$~\cite{Avgoustidis:2011ax,Pourtsidou:2010gu}.
        \\*
        \\*
        \indent In this paper we use the Planck2015 public data~\cite{Aghanim:2015xee} to perform the first full Markov chain Monte Carlo (MCMC) analysis of $\Lambda$ cold dark mater (CDM) models with cosmic string or superstring networks. For ``ordinary" cosmic string networks we work in the unconnected segment model (USM) framework and utilise our analytic method~\cite{Avgoustidis:2012gb} for fast computation of the string unequal-time correlator (UETC). This is used as a source to compute CMB anisotropies and hence obtain joint constraints on $\Lambda$CDM and the string network parameters, including the  tension $G\mu$, the loop chopping efficiency $c_{\rm r}$, and the wiggliness parameter $\alpha$. In the case of cosmic superstring networks we extend our method to deal with multiple network components. The UETC approach is efficient, meaning we can compute the superstring spectrum in much less time than previous codes and obtain joint constraints on the fundamental string tension $G\mu_{\rm F}$, the string coupling constant $g_{\rm s}$, the self-interaction coefficient $c_{\rm s}$, and the parameter $w$ of \cite{Pourtsidou:2010gu}, quantifying the volume of compact extra dimensions.
        \\*
        \\*
        \indent In Sec.~\ref{sec:UETC} we describe the  UETC formalism applied to evolving Nambu-Goto string networks. In Sec.~\ref{sec:superstrings} we summarise our modelling of cosmic superstrings and the adaptation of our UETC method to these multistring component networks. In Sec.~\ref{sec:constraints} we present the results of our MCMC analysis for cosmic string and superstring networks using Planck2015 CMB data. Our constraints on string network parameters and possible future directions are discussed in Sec.~\ref{sec:discussion}.
                
\section{Unequal-Time Correlator}
        \label{sec:UETC}
        
            \indent Unlike passive inflationary perturbations which are set as initial conditions, metric perturbations from cosmic string networks are actively sourced at all times. To compute the string spectra the components of the string network's energy-momentum tensor must be used as sources in the linearised Einstein-Boltzmann equations. The relevant quantity to calculate is UETC, whose dominant eigenmodes, found by diagonalising, can be used as source functions, with each individual mode being coherent~\cite{Pen:1997ae}. The UETC
            \vskip1.5px
  \begin{equation}
    \langle\Theta_{\mu\nu}({\bf k},\tau)\Theta^*_{\alpha\beta}({\bf k},\tau')\rangle\equiv\mathcal{C}_{\mu\nu,\alpha\beta}({\bf k},\tau,\tau')
    \end{equation}
        \vskip1.5px
    \noindent determines all the two-point correlation functions such as the CMB temperature $C_\ell$ and matter power spectra $P(k)$, defined as in \cite{Brandenberger:2000}. $\Theta_{\mu\nu}({\bf k},\tau)$ is the string energy-momentum tensor defined below.

        \subsection{String energy-momentum tensor}
            Nambu-Goto strings are one-dimensional defects in the zero-width limit. They provide a good description for long cosmic strings, whose correlation length is many orders of magnitude larger than their width, at least away from string intersections. A string moving in spacetime spans a two-dimensional surface, the worldsheet $x^\mu(\sigma^a)$, where the indices $\mu=0,1,2,3$ label spacetime coordinates and $a=0,1$ are the indices of coordinates on the worldsheet~\cite{Hindmarsh:1994,Nambu:1970}. The worldsheet action is reparametrisation invariant and a gauge can be chosen by imposing two conditions on the spacetime coordinates $x^\mu$ as functions of $\sigma^a$. In a Friedmann-Robertson-Walker (FRW) background, a useful choice of gauge is such that $\sigma^0=\tau$, the conformal time, and ${\bf x}'\cdot{\dot{\bf x}}=0$, where $\dot{~}\equiv\partial/\partial\tau$ and $'\equiv\partial/\partial\sigma$, relabelling $\sigma^1$, which in this gauge is a spacelike worldsheet coordinate, as $\sigma$. In this gauge the Nambu-Goto string energy-momentum tensor is
            \begin{widetext}
            \begin{equation}
                \Theta^{\mu\nu}(y)=\frac{1}{\sqrt{-g}}\int d\tau d\sigma\bigg[U\sqrt{-\frac{{\bf x}'^2}{\dot{{\bf x}}^2}}\dot{x}^\mu\dot{x}^\nu-T \sqrt{-\frac{\dot{\bf{x}}^2}{{\bf x}'^2}}x'^\mu x'^\nu\bigg]\delta^{(4)}(y-x(\tau,\sigma))\,.\label{eqn:emt}
            \end{equation}
            \end{widetext}
            Here, $U$ is the string energy per unit length and $T$ is the string tension. For Nambu-Goto strings on arbitrarily small scales, Lorentz invariance requires that $T=U=\mu$. However, if we coarse grain the string, then the integrated effect of small-scale structure is to make the effective tension smaller than the energy density. We can then include the effect of small-scale wiggles on the string via a ``string wiggliness" parameter $\alpha$, such that
            \begin{equation}
                U=\alpha\mu~{\rm and}~T=\frac{\mu}{\alpha}\,,
            \end{equation}
            satisfying $UT=\mu^2$.
            \\*
            \\*
            \indent The Fourier transform of the 00-component of the energy-momentum tensor of a representative string segment in a network is
            \begin{align}
                \Theta_{00}(\tau,{\bf k},\chi) =& \frac{\mu\alpha}{\sqrt{1-v^2}}\frac{\sin({\bf k}\cdot{ \hat{\bf X}}\xi\tau/2)}{{\bf k}\cdot\hat{\bf X}/2}\nonumber\\
                &\phantom{space}\times\cos\left({\bf k}\cdot{\bf x}_0+{\bf k}\cdot{\dot{\hat{\bf X}}}v\tau\right)\,,\label{eq:enmote}   
            \end{align}
            where $v$ and $\xi$ are the string network velocity and comoving correlation length, defined in Sec.~\ref{sec:VOS} below, and ${\bf x}_0$ is the position of the string end point. The string segment is parametrised by 
            \begin{equation}
            {\bf x}(\sigma,\tau)={\bf x}_0+\sigma{\hat{\bf X}}+v\tau{\dot{\hat{\bf X}}}\,,
            \end{equation}
            with the string orientations and velocity orientations 
            \begin{align}
                {\hat{\bf X}}=&\begin{pmatrix}\sin\theta\cos\phi\\\sin\theta\sin\phi\\\cos\theta\end{pmatrix}\,,\\
                {\dot{\hat{\bf X}}}=&\begin{pmatrix}\cos\theta\cos\phi\cos\psi-\sin\phi\sin\psi\\\cos\theta\sin\phi\cos\psi+\cos\phi\sin\psi\\-\sin\theta\cos\psi\end{pmatrix}\,.
            \end{align}
            ${\dot{\hat{\bf X}}}$ is transverse to ${\hat{\bf X}}$ such that ${\hat{\bf X}}\cdot{\dot{\hat{\bf X}}}=0$. Note that the position of the string end point appears only through a phase in the cosine factor in Eq.~(\ref{eq:enmote}), which we will denote as $\chi\equiv {\bf k}\cdot{\bf x}_0$. The other components of the string energy-momentum tensor are given by
            \begin{equation}\label{eq:enmoteii}
                \Theta_{ij}=\bigg(v^2{\dot{\hat{\bf X}}}_i{\dot{\hat{\bf X}}}_j-\frac{1-v^2}{\alpha^2}{\hat{\bf X}}_i{\hat{\bf X}}_j\bigg)\Theta_{00}\,,
            \end{equation}
            with $i,j=1,2,3$. Choosing coordinates so that ${\bf k}$ lies along the $\hat{k}_3$ axis, the scalar, vector and tensor anisotropic stresses are given by
            \begin{align}
                \Theta^{\rm S}=&\phantom{+}\frac{1}{2}(2\Theta_{33}-\Theta_{11}-\Theta_{22})\,,\label{eq:thetas}\\
                \Theta^{\rm V}=&\phantom{\frac{1}{2}}\Theta_{13}\,,\label{eq:thetav}\\
                \Theta^{\rm T}=&\phantom{\frac{1}{2}}\Theta_{12}\,.\label{eq:thetat}
            \end{align}
            
        \subsection{Velocity dependent one-scale model}
        \label{sec:VOS}

            The velocity one-scale model (VOS) equations dictate the values of the string network correlation length $L$, and the average velocity $v$, of string segments in the network~\cite{Kibble:1984}. The correlation length $L$ is the average length of string segments which, for scaling networks (that have a random walk structure), is also equal to the average string separation. The network velocity $v$, is the root-mean-square (rms) velocity of these correlation-length-sized string segments averaged over all (shorter) length scales. The macroscopic evolution equations for these network parameters can be derived from the Nambu-Goto action by applying a statistical averaging procedure over the string worldsheet~\cite{Martins:1995,Martins:1996,Martins:2000}. Expressed in terms of the physical time $t$, they read
            \begin{align}
            \dot{L}=&\phantom{+}(1+v^2)L\frac{\dot{a}}{a}+\frac{c_{\rm r}v}{2}\,,\\
            \dot{v}=&\phantom{+}(1-v^2)\bigg(\frac{\tilde{k}}{L}-2v\frac{\dot{a}}{a}\bigg)\,, 
            \end{align}
            where $a(t)$ is the scale factor, $\dot{a}(t)/a(t)$ is the Hubble function and from now on $\dot{~}\equiv d/dt$, unlike in Eq.~(\ref{eqn:emt}). The loop chopping efficiency parameter, $c_{\rm r}$, quantifies the energy loss due to loop production and $\tilde{k}$ provides a phenomenological description of the small-scale structure on the string, which, for relativistic strings, is given by
            \begin{equation}
                \tilde{k}=\frac{2\sqrt{2}}{\pi}\bigg(\frac{1-8v^6}{1+8v^6}\bigg)\,.\label{kofv}
            \end{equation}
            The correlation length can be written in comoving units as $\xi\tau=L/a$. The VOS equations in comoving units are
            \begin{align}
            \xi'=&\phantom{+}\frac{1}{\tau}\bigg(v^2\xi\tau\frac{a'}{a}-\xi+\frac{c_{\rm r}v}{2}\bigg),\label{VOSxi_comoving}\\
            v'=&\phantom{+}(1-v^2)\bigg(\frac{\tilde{k}}{\xi\tau}-2v\frac{a'}{a}\bigg)\,, \label{VOSv_comoving}
            \end{align}
            where now $'\equiv d/d\tau$, unlike in Eq.~(\ref{eqn:emt}).
            For fixed expansion rate the scaling solutions, found by the requirement $\xi'=0$ and $v'=0$, read
            \begin{align}
            \xi=&\phantom{+}\sqrt{\frac{\tilde{k}(\tilde{k}+c_{\rm r})(1-\beta)}{4\beta}}\,,\\
            v=&\phantom{+}\sqrt{\frac{\tilde{k}(1-\beta)}{\beta(\tilde{k}+c_{\rm r})}} \label{scaling_sol}\,,
            \end{align}
            where $\beta$ is the physical time FRW expansion exponent $a(t)\propto t^\beta$ and is equal to $1/2$ and $2/3$ in the radiation and matter eras respectively. 
            Note in the scaling solutions of (\ref{scaling_sol}) the implicit velocity dependence of $\tilde k$ through Eq.~(\ref{kofv}). Earlier implementations of the cosmic defect CMB code {\ttfamily CMBACT}~\cite{Pogosian:1999np} used two sets of values for the loop chopping efficiency and the parameter $\tilde k$ in the scaling solutions (\ref{scaling_sol}) for the radiation and matter eras. These values were then interpolated for the transition between the radiation and matter eras. However, in the latest implementation of the VOS equations in {\ttfamily CMBACT4}~\cite{Pogosian:2014}, the velocity dependence of $\tilde k$ is explicitly used and the loop chopping efficiency is kept constant throughout both epochs~\cite{Martins:2000}. Here, we also adopt this approach: at any particular $\tau$, the values of $\xi$ and $v$ found using the VOS equations [(\ref{VOSxi_comoving}) and (\ref{VOSv_comoving})] are used for calculating the UETC, keeping  $c_{\rm r}$ constant throughout and explicitly accounting for the velocity dependence ~(\ref{kofv}) of $\tilde k$. In earlier versions of {\ttfamily CMBACT} the wiggliness $\alpha$, was also an evolving parameter, but it is now kept constant in {\ttfamily CMBACT4}, which is the approach we take here. 
             \begin{figure}
 \centering
 \includegraphics{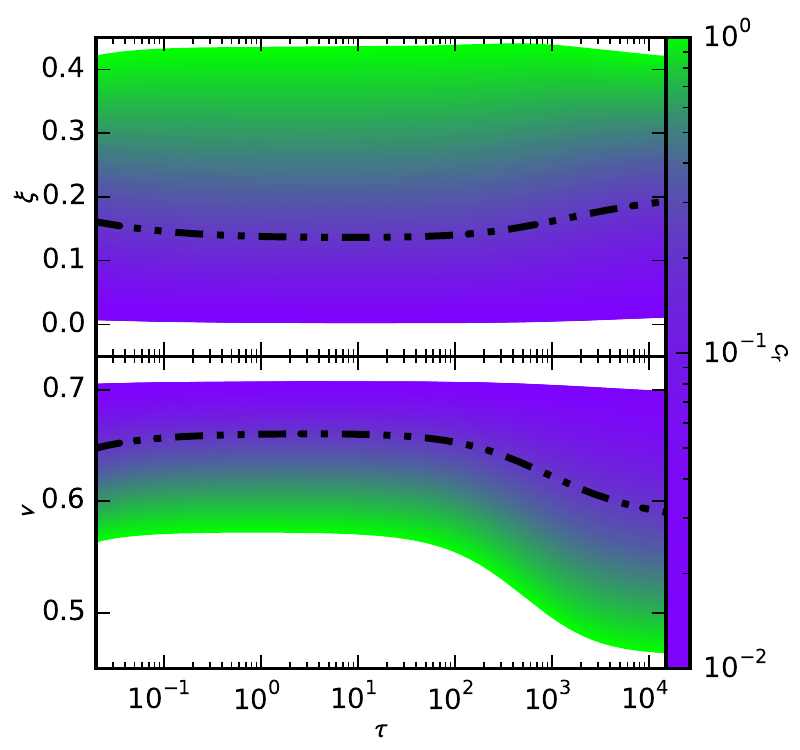}
 \vskip-10pt
 \caption{The evolution of the velocity $v$, and correlation length $\xi$, for a range of $c_{\rm r}=[10^{-2},1.0]$. The black dot-dot-dash line indicates the correlation lengths and velocities obtained when $c_{\rm r}=0.23$. The greener area (lighter in black and white) of the plot indicates larger values of $c_{\rm r}$ whilst the more purple region (darker in black and white) shows a smaller $c_{\rm r}$.}
 \label{fig:ev}
 \vskip-15pt
 \end{figure}
            The evolution of the network parameters can be seen for a range of $c_{\rm r}$ in Fig.~\ref{fig:ev} showing that a wide range of correlation lengths and velocities are available. Detailed comparison of the VOS model with Nambu-Goto simulations of ordinary string networks (i.e. single string type with unit intercommuting probability~\cite{Shell_Recon}) determine the loop chopping efficiency to $c_{\rm r}=0.23 \pm 0.04$~\cite{Martins:2000}, corresponding to the black dot-dot-dashed curves in Fig.~\ref{fig:ev}. Models of cosmic superstrings generally have suppressed intercommutation probabilities~\cite{Jones:2003da,Jackson:2004zg,Hanany:2005bc,Jackson:2007hn}, which effectively reduces $c_{\rm r}$, and so they correspond to the purple region in the figure. Such networks have relativistic rms velocities $v\sim 1/\sqrt{2}$ and correlation lengths much smaller than the horizon, corresponding to a much higher string number density compared to ordinary string networks. However, they also have smaller string tension so their overall effect on the CMB can be small, consistent with the data.\\*
		\\*
		\indent It should be noted that the rms network velocity used in the VOS model arises from a worldsheet average and is thus integrated over all (short) length scales. Therefore, it provides an accurate measure of the energy stored in a wiggly string segment, but does not explicitly correspond to (and in fact is expected to be larger than) the coherent velocity on correlation-length scales. Indeed, numerical simulations of Nambu-Goto strings reveal a network velocity distribution with larger velocities at short scales, implying that the rms velocity is dominated by relativistic speeds at short distances. On length scales of order the correlation length, coherent velocities as low as $v_{\rm coh}\simeq 0.2$ have been reported \cite{BennetBuchet,AllenShellard,Martins_fractal,Vieira_SSS}. Other network velocity measures (again containing information from a range of length scales) in both Nambu-Goto and Abelian-Higgs string simulations also tend to be lower than the VOS rms velocity, with velocities in the Abelian-Higgs model $v_{AH}\simeq 0.5$, significantly slower than in Nambu-Goto simulations~\cite{Hindmarsh:2008dw,Bevis:2010gj,BlancoPillado:2011dq}.  For further discussion about the impact of string velocities on the UETC and the string power spectrum see the end of Sec.~\ref{sec:comp}.     

        \subsection{Unconnected segment model}
            
		Simulations of evolving string networks are numerically very expensive. Strings decay as $1/(\xi\tau)^{3}$, eventually reaching a scaling solution ($\xi=\rm constant$) with a number density of tens to hundreds of strings per horizon volume. At early times, the box contains a huge number of strings whose dynamics and interactions have to be tracked at each time step. The USM~\cite{Albrecht:1997mz,Pogosian:1999np} dramatically reduces the required computational resources by approximating the string network as a collection of correlation-length-sized segments, with the time evolution of the correlation length and segment velocity described by the VOS equations. Moreover, the model consolidates these string segments by collecting all strings that decay between any two times, and so fewer strings will need to be tracked. The number of strings that decay between any two conformal times in a volume $V$ is
            \begin{equation}
                N_{\rm d}(\tau_i)=V[n(\tau_{i-1})-n(\tau_i)]\,,
            \end{equation}
            where $n(\tau)$ is the number density of strings at conformal time $\tau$, given by $n(\tau)=C(\tau)/(\xi\tau)^3$. In {\ttfamily CMBACT}, the factor $C(\tau)$ is chosen so as to keep the number of strings at any time proportional to $1/(\xi\tau)^3$. The energy-momentum tensor for the string network is then given by the sum over the total number of consolidated string segments $K$ , with a factor accounting for string decay
            \begin{equation}
            \Theta_{\mu\nu}=\sum_{i=1}^K\sqrt{N_{\rm d}(\tau_i)}\Theta_{\mu\nu}^iT^{\rm off}(\tau,\tau_i,L_{\rm f}).
            \end{equation}
            The string decay factor $T^{\rm off}(\tau,\tau_i,L_{\rm f})$ is a function interpolating between 1 and 0 and is responsible for turning off the contribution of the $i^{\rm th}$ consolidated segment after the time it has decayed.  Its steepness is controlled by a string decay parameter, $0<L_{\rm f}\leq 1$, as follows:
            \begin{equation}
                T^{\rm off}(\tau,\tau_i,L_{\rm f})=\left\{\begin{array}{cl}1&\tau<L_{\rm f}\tau_i\\1/2+1/4(y^3-3y)&L_{\rm f}\tau_i<\tau<\tau_i\\0&\tau_i<\tau\end{array}\right.\label{eq:Toff}
            \end{equation}
            where
            \begin{equation}
            y=\frac{2\ln(L_{\rm f}\tau_i/\tau)}{\ln(L_{\rm f})}-1\,.
            \end{equation}
		Thus, in the limit $L_{\rm f}\!\to\!1$ the string decay factor $T^{\rm off}(\tau,\tau_i,L_{\rm f})$ approaches a Heaviside function, sharply switching off the contribution of the $i^{\rm th}$ consolidated segment to the network energy-momentum tensor for times $\tau>\tau_i$.

            \subsubsection*{The $L_{\rm f}$ Parameter}
            
\begin{figure*}
                    \centering
                    \includegraphics{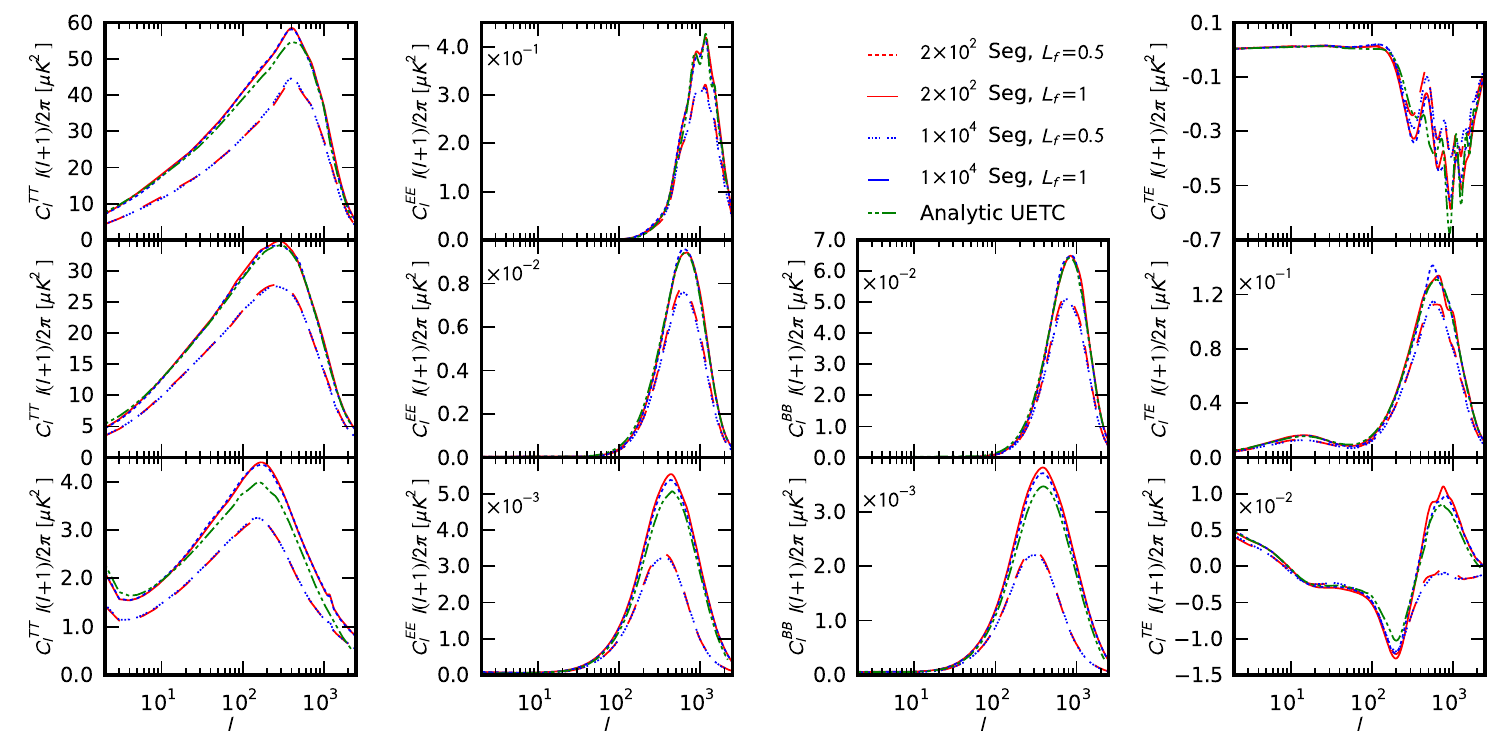}
                    \vskip-15pt
                    \caption{$C_\ell$ obtained from the string realisation code {\ttfamily CMBACT4} with 200 and 10000 consolidated string segments for 2000 string realisations between the red solid and dashed lines and blue dotted and dotted-segment lines respectively. The solid red and dotted blue lines at the top of each band indicate a value of $L_{\rm f}=1$ for 200 and 10000 segments, while the red dotted and blue dotted-segment lines show $L_{\rm f}=0.5$. The top, middle and bottom rows show the scalar, vector and tensor $C_\ell$ modes respectively. The first column contains the temperature (TT) $C_\ell$, the second column has the EE-mode contribution, BB-modes are in the third and the TE cross-correlation is in the final column. We also plot the corresponding spectra derived from our analytic USM method, shown in green dot-dot-dashed lines.}
                    \label{fig:cls_cmbact}
                    \vskip-15pt
                \end{figure*}

        Since the number of consolidated segments also sets the number of decay epochs, a finite number of consolidated segments leads to discrete steps in the number density of strings. The string decay parameter $L_{\rm f}$ was introduced to allow a fraction of the consolidated strings to decay before the end of their respective decay epoch, thus making the number density evolution smoother. The function $C(\tau)$ was also introduced to ensure that the number of strings at any conformal time $\tau$ is kept proportional to $(\xi\tau)^{-3}$. However, one consequence of $L_{\rm f} < 1$ is that it is possible that $L_{\rm f}\tau_{i+1} < \tau_i $, meaning strings can start to decay earlier than their respective epoch and the number density is systematically lower.\\*  
        \\*
 \indent In the  {\ttfamily CMBACT4} implementation  we have found that changing the number of consolidated segments from 200 to 10000  has very little impact on the string spectra,  as shown in  Fig.~\ref{fig:cls_cmbact}. However, the amplitude of the $C_\ell$  {\em is} dependent on the value of $L_{\rm f}$. The change is scale dependent, but can be as much as $30\%$, for example, near the peak of the scalar temperature signal. Previous analyses which have used the results from {\ttfamily CMBACT} have overlooked this dependence. Although not entirely degenerate with the amplitude of $C_{\ell}$, which scales proportional to $(G\mu)^2$, it will clearly have some effect on the inferred values of $G\mu$ from the USM. We compare this to our approach in the following section.

            \subsubsection*{Infinite Consolidated String Segments}
            
            We are able to accommodate a large number of segments analytically. As discussed in \cite{Avgoustidis:2012gb}, the scaling factor, that weights the UETC taking into account string decay, has a particularly simple form when the number of consolidated string segments tends to infinity, $L_{\rm f}\to 1$ and $C(\tau)\to 1$. This is
                \begin{align}
                    f(\tau_1,\tau_2,\xi(\tau_1),\xi(\tau_2))=&\phantom{+}\sum_{i=1}^KN_{\rm d}(\tau_i)T^{\rm off}(\tau_1,\tau_i,L_{\rm f})\nonumber\\&\phantom{space}\times T^{\rm off}(\tau_2,\tau_i,L_{\rm f}),\nonumber\\
                    =&\phantom{+}(\xi({\rm Max}[\tau_1,\tau_2]){\rm Max}[\tau_1,\tau_2])^{-3}.\nonumber\\
		    =&\phantom{+}f\big(\tau_{\rm Max},\xi(\tau_{\rm Max})\big)\label{eq:scal}
                \end{align}
                An analytic expression for the scaling factor can also be found for an arbitrary $L_{\rm f}$ using the form of $T_{\rm off}$ quoted in Eq.~(\ref{eq:Toff}). However, it seems natural to consider only the case $L_{\rm f}=1$ when the number of consolidated string segments is very large. In the infinite limit  the segments will decay at an infinite number of epochs which are infinitesimally separated, a continuous limit in which the string number density is smooth. We have shown that the number density scales according to $(\xi\tau)^{-3}$  with our approach. While infinite consolidated segments may seem unphysical, it is just a limit used to obtain the correct scaling relation. We obtain very similar results to {\ttfamily CMBACT4}  when using between 200 to 10000 segments with    $L_{\rm f}=1$. The question of whether the observed resulting modification of scaling  from early string decay obtained when $L_{\rm f}<1$ is physical or not requires investigation. Since we take $C(\tau)=1$ we avoid considering different scaling behaviour. Ultimately, the USM is a simplified model which aims to match the UETC from simulations by adjusting the network parameters. Overall it has been shown to match Nambu-Goto simulations well~\cite{Lazanu:2014xxa}.  However, due to the correlation between the inferred values for $G\mu$ for a given $L_{\rm f}$, this issue should be considered more closely.\\*
     \\*
     \indent Since the number density scales according to $(\xi\tau)^{-3}$ using our approach, we believe this to be reasonable and will adopt this for the comparison to data.
     
            \subsection{Analytic calculation of the unequal-time correlator}
            
\begin{figure}
	    \centering
	        \includegraphics{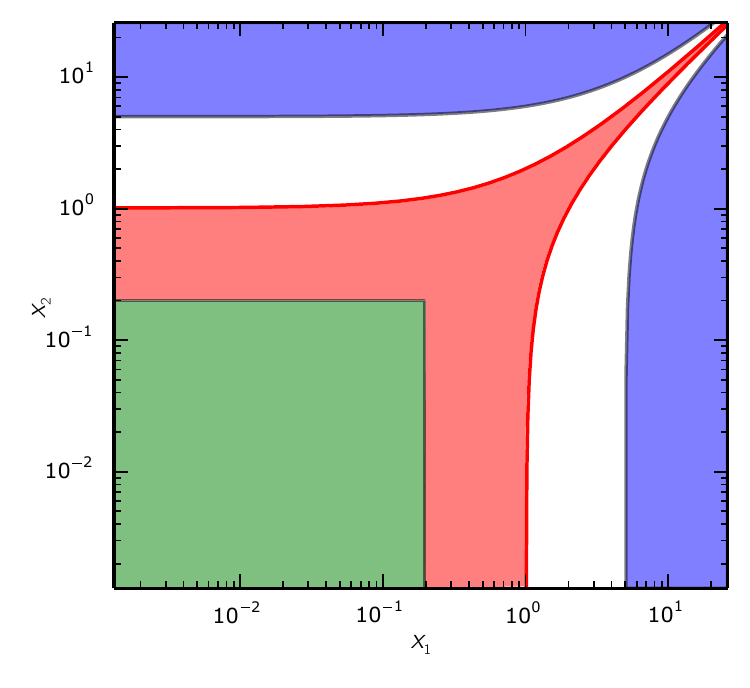}
	        \vskip-15pt
		\caption{The regions of $x=k\tau\xi$ covered by analytic approximations. In green is the region where $x_1\ll1$ and $x_2\ll1$, red where $|\log{x_1}-\log{x_2}|<{\rm \epsilon}$, and blue where $|x_1-x_2|\gg1$. In the code the $x_{1,2}\ll1$ region is set for $x_{1,2}<0.2$, $\epsilon=0.001$ for $x_1\approx x_2$, and $|x_1-x_2|>10$ for $x_{1,2}\gg x_{2,1}$.}
	        \label{fig:cover}
	        \vskip-15pt
	 \end{figure}

                The UETC can be computed analytically \cite{Avgoustidis:2012gb} by integrating over all string configurations (orientations and positions) in the network. For the two-point correlator between $\Theta(\tau_1,{\bf k}_1,\chi_1)$ and $\Theta(\tau_2,{\bf k}_2,\chi_2)$ translational invariance implies ${\bf k}_1=-{\bf k}_2={\bf k}$ and so $\chi_1=-\chi_2=\chi$. Considering that, due to Eqs.~(\ref{eq:enmote}) and ~(\ref{eq:enmoteii}), $\Theta(\tau,{\bf k},\chi)$ is a symmetric function of ${\bf k}$ the integral is
                \begin{widetext}
                \vskip-0.4cm
                \begin{align}
                    \langle\Theta(\tau_1,{\bf k})\Theta(\tau_2,{\bf k})\rangle=\frac{2f(\tau_{\rm Max},\xi(\tau_{\rm Max}))}{16\pi^3}\int_0^{2\pi}d\phi\int_0^{2\pi}d\psi\int_0^\pi\sin\theta d\theta\int_0^{2\pi}d\chi\Theta(\tau_1,{\bf k},\chi)\Theta(\tau_2,{\bf k},\chi).
                \end{align}
                \end{widetext}                
                Without loss of generality ${\bf k}$ can be chosen to lie along the $k_3$-axis, such that ${\bf k}=k\hat{k}_3$. $\Theta$ here represents each of $\Theta_{00}$, $\Theta^{\rm S}$, $\Theta^{\rm V}$ and $\Theta^{\rm T}$ of Eqs.~(\ref{eq:thetas}-\ref{eq:thetat}). The $\phi$, $\psi$, and $\chi$ integrals can be done analytically in this case, leaving only the $\theta$ integral in terms of Bessel functions. The UETC can then be written as the sum over six integral identities,
                \begin{widetext}
                \vskip-0.4cm
                \begin{equation}
                \langle\Theta(\tau_1,k)\Theta(\tau_2,k)\rangle=\frac{f(\tau_{\rm Max},\xi(\tau_{\rm Max}))\mu^2}{k^2\sqrt{1-v(\tau_1)^2}\sqrt{1-v(\tau_2)^2}}\sum_{i=1}^6A_i[I_i(x_-,\varrho)-I_i(x_+,\varrho)],\label{UETCum}
                \end{equation}
                \end{widetext}
                where $\varrho=k|v(\tau_1)\tau_1-v(\tau_2)\tau_2|$ and $x_{\pm}=(x_1\pm x_2)/2$ with $x_{1,2}=k\xi(\tau_{1,2})\tau_{1,2}$. Here $x_{1,2}$ means $x_1$ or $x_2$ respectively. This extends the corresponding result of \cite{Avgoustidis:2012gb} in that $\xi$ and $v$ are now functions of $\tau$ instead of being kept constant. This means that the expressions of the amplitudes $A_i$, presented in Table~\ref{tab:amplitudes}, are now time-dependent. The integral identities (shown in Table~\ref{tab:integrals}) remain the same. 
                It should be noted that $I_1(x,\varrho)$ and $I_4(x,\varrho)$ diverge but the combination $I_{1,4}(x_-,\varrho)-I_{1,4}(x_+,\varrho)$ is regular and, in the limit where $x_{1,2}\gg x_{2,1}$, has an analytic approximation given by
                \begin{align}
                    I_{1}(x_-,\varrho)-I_{1}(x_+,\varrho)=&\phantom{+}\frac{\pi x_{1,2}}{2}J_0(\varrho),
                \end{align}
                \begin{align}
                    I_{4}(x_-,\varrho)-I_{4}(x_+,\varrho)=&\phantom{+}\frac{\pi x_{1,2}}{2\varrho}J_1(\varrho).
                \end{align}
                In the small $x_{1,2}$ limit, the UETC can be written as 
                \begin{equation}
                    \langle\Theta(\tau_1,k)\Theta(\tau_2,k)\rangle=\frac{f(\tau_{\rm Max},\xi(\tau_{\rm Max}))\mu^2}{k^2\sqrt{1-v(\tau_1)^2}\sqrt{1-v(\tau_2)^2}}B,
                \end{equation}
                and at equal times, when $x_1=x_2=x$ and $\varrho=0$, the equal-time correlator is given by
                \begin{equation}
                    \langle\Theta(\tau,k)\Theta(\tau,k)\rangle=\frac{f(\tau,\xi(\tau))\mu^2}{k^2(1-v(\tau)^2)}C.
                \end{equation}
                The forms of $B$ and $C$ are similar to \cite{Avgoustidis:2012gb} but again depend on the values of $v$ and $\xi$ at $\tau_1$ and $\tau_2$. These coefficients have also been included in Table~\ref{tab:amplitudes}. Thanks to these analytic approximations, computational times can be greatly reduced compared to the case where the integral identities $I_i$ are used for computation over the whole range of $k\tau_1$, $k\tau_2$. The regions where these approximations are valid are shown in Fig.~\ref{fig:cover}, only the white region is computationally intensive. It should be noted that, because $\xi$ is a function of time, the shape of the approximated regions in Fig.~\ref{fig:cover} changes for different values of $k$ and so we must consider a large number of $k$-modes when computing the UETC. This is in contrast to ~\cite{Avgoustidis:2012gb}, where the approximation of constant $\xi$ and $v$ meant that the UETC was only a function of the combinations $k\tau_1$ and $k\tau_2$.   

                \subsubsection*{Negative values of the UETC}
    
\begin{figure}
        \centering
        \vskip-15pt
        \includegraphics{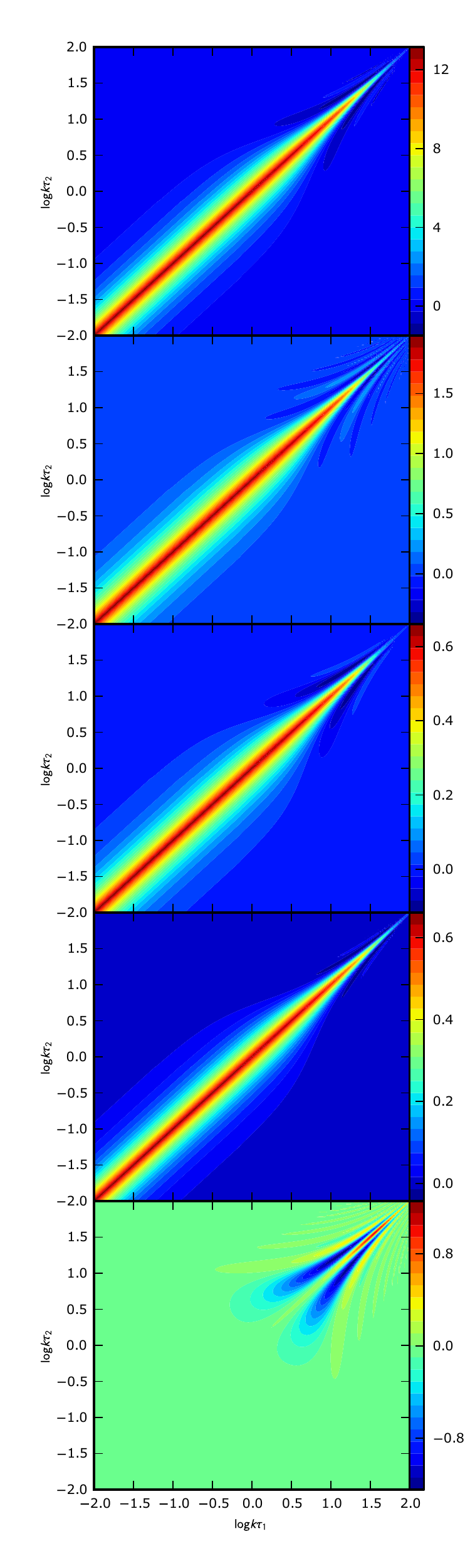}
        \vskip-20pt
        \caption{The UETC calculated at $k=0.05h/$Mpc. The plots show 00-component followed by the scalar, vector, and tensor anisotropic stresses. (Bottom panel) The cross-correlation between the energy-density and the scalar anisotropic stress.}
	     \label{fig:uetc}   
    \end{figure}

\begin{figure}
        \centering
        \vskip-15pt
        \includegraphics{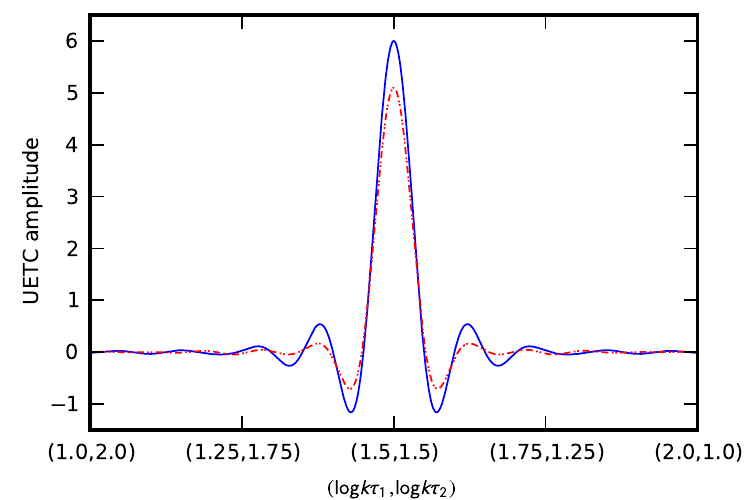}
        \caption{Profile of the UETC across the diagonal in the oscillatory region with a large $k\tau_1\approx k\tau_2$. The solid blue line shows the amplitude of the UETC using the value of the velocity and the correlation length from the VOS equations whilst the red dot-dot-dash line has Gaussian distributed velocities and correlation lengths about the VOS values.}
	\label{fig:dip}   
    \end{figure}

                It has been noted in \cite{Regan:2014vha} that there are negative regions in the string UETC calculated analytically through our formalism, which do not appear in the Gaussian model for the string UETC used in \cite{Regan:2014vha}. These can be seen in Fig.~\ref{fig:uetc}.\\*
                \\*
\indent There are two distinct types of regions with negative values of our UETC. First, regions with small $k\tau_1$ and large $k\tau_2$ (and vice versa), corresponding to the top left and bottom right corners of Fig.~\ref{fig:cover} or Fig.~\ref{fig:uetc}: in these regions the UETC should be zero, but small negative (and positive) values can arise from the finite order truncation of the Bessel series expansions of $I_1(x_{\pm},\rho)$ and $I_4(x_{\pm},\rho)$ in Eq.~(\ref{UETCum}). These values are spurious and can be thought of as noise arising from the truncation. The order of truncation must then be chosen such that this noise is at a tolerable level.\\*
\\*
\indent Second, in the regions off the diagonal with a large $k\tau_1\approx k\tau_2$ (corresponding to the top right corner of Fig.~\ref{fig:cover} or Fig.~\ref{fig:uetc}) there is a ringing pattern with successive positive and negative peaks that decay as we move away from the diagonal. These oscillatory patterns are a consequence of causality~\cite{Turok:1996ud,Durrer:1997ep,Albrecht:1997mz}, built into the USM: as the correlator must vanish at superhorizon scales (in fact in the USM it vanishes at scales larger than the correlation length, which is smaller than the horizon), this introduces a sharp edge in physical space that becomes oscillatory in Fourier space. This oscillatory pattern therefore has a clear physical origin, but in the USM it is somewhat artificially enhanced due to the fact that the model assumes all string segments have the same length. If segments are instead given a length distribution peaking at the network correlation length, the sharp edge is smoothed and the oscillatory pattern gets suppressed. Furthermore, considering a segment velocity distribution peaking near the network rms velocity again suppresses these oscillations. The Gaussian model assumes a wide Gaussian distribution of string lengths (but also assigns non-zero values to the correlator at superhorizon scales), so this causal oscillatory feature is absent from the UETC in that model.\\*
\\*
\indent The suppression of oscillations in the UETC can be seen in Fig.~\ref{fig:dip} where the blue solid line shows the profile of the UETC across the diagonal as calculated using the velocity and correlation lengths from the VOS. The red dot-dot-dash line is the same profile when a Gaussian distributed sample of velocities and correlation lengths, peaking on the VOS values, is chosen. The oscillatory features are mostly washed out but the first trough remains a prominent feature. The off-diagonal dip in the correlation functions that we find after considering a range of segment lengths and velocities has also been observed in Abelian-Higgs simulations~\cite{Bevis:2010gj}. It may also be related to the velocity anti-correlation observed in Nambu-Goto simulations on correlation-length scales and can be attributed to string intercommutations~\cite{Martins_fractal}.   

            \subsection{Eigenmode decomposition}
            \label{sec:eig}
            
\begin{figure*}
                     \centering
                     \includegraphics{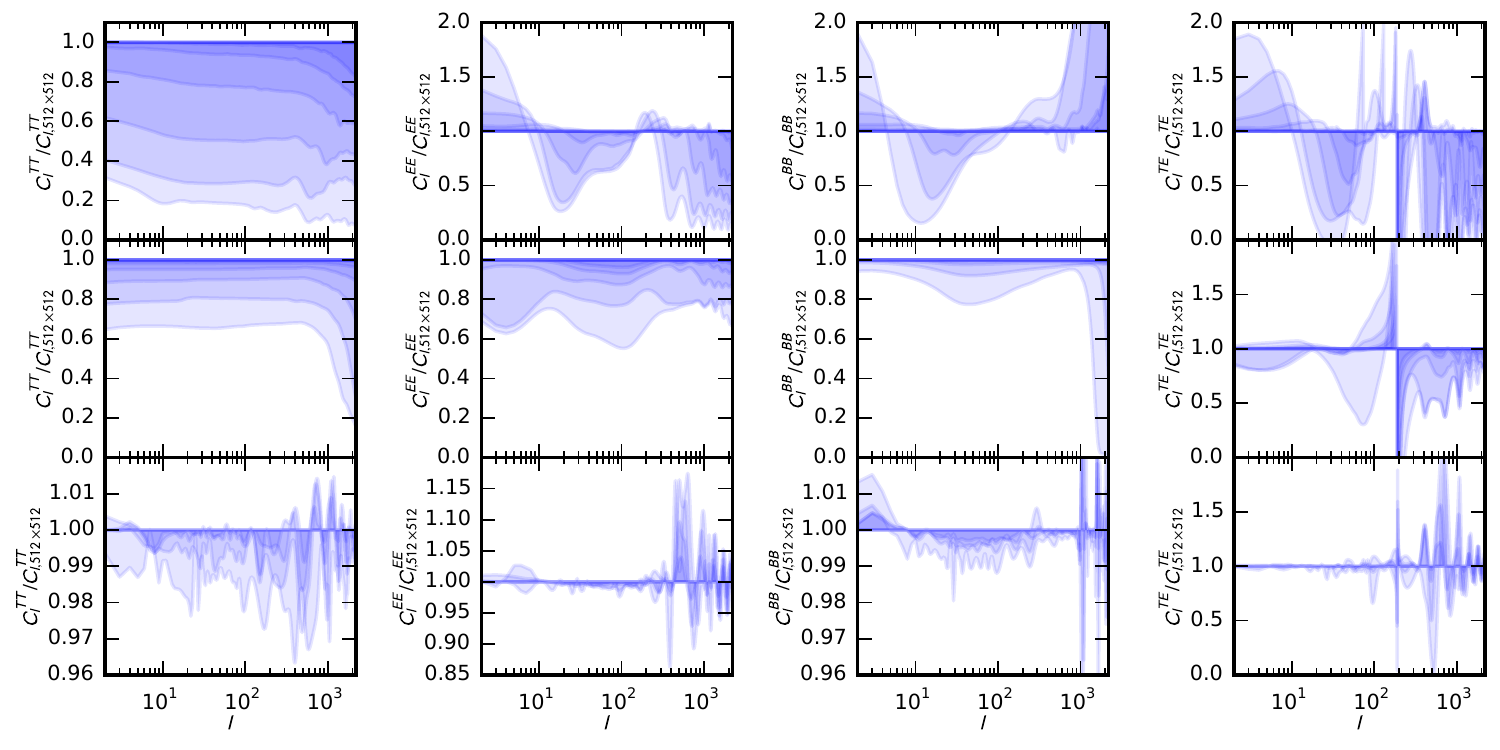}
                     \vskip-15pt
 \caption{The ratio of the $C_\ell$ calculated using a UETC with a $512\times512$ grid with all the eigenmodes and (top row) other grid resolutions, the lightest shaded region with an $8\times8$ grid and the darkest a $256\times256$ grid; (middle row) when fewer eigenmodes are included, eight eigenmodes for the lightest shaded region and 256 for the darkest; (bottom row) when the {\ttfamily accuracy\_boost} setting of {\ttfamily CAMB} is decreased, reducing the number of $k$ values.}
                     \label{fig:cls}
                     \vskip-15pt
                  \end{figure*}

                The UETC is generally rescaled by a factor of $\sqrt{\tau_1\tau_2}$, which, for $\xi$ and $v$ constant, makes it a function of $k\tau_1$ and $k\tau_2$ only. This is not true in the present case because now we are tracking the time-dependence of $\xi$ and $v$, so the UETC depends separately on $k$, $\tau_1$, and $\tau_2$.  However, it is still useful to introduce this rescaling in order to facilitate direct comparison of the UETC with previous results. This rescaled UETC can then be discretised onto a logarithmic grid in $k\tau_1$ and $k\tau_2$ with $n\times n$ grid points and then diagonalised giving the eigenvectors and eigenvalues~\cite{Pen:1997ae}
                \begin{align}
                &(k^2\tau_1\tau_2)^\gamma\sqrt{\tau_1\tau_2}\langle\Theta(\tau_1,k)\Theta(\tau_2,k)\rangle=\nonumber\\
                &\phantom{spacespacespace}\sum_{i=1}^N\lambda_iu_i(k\tau_1)\otimes u_i(k\tau_2).
                \end{align}
                Due to the explicit dependence on $k$, this diagonalisation procedure has to be repeated for a large number of $k$-modes, and the eigenvalues are $k$-dependent. This significantly increases the computation time compared to \cite{Avgoustidis:2012gb}. The extra factor $(k^2\tau_1\tau_2)^\gamma$ is used for more efficient reconstruction of the UETC when the eigenmodes are truncated below $n$.  The choice $\gamma=0.25$ gives the best reconstruction on scales that give the dominant contribution to the CMB anisotropies.\\* 
                \\*
                \indent There is no correlation between the scalar, vector, and tensor modes, so the vector and tensor UETC can be diagonalised independently. However, the density $\Theta_{00}$ and the scalar anisotropic stress $\Theta^{\rm S}$ are correlated. The diagonalisation is done over a $2n\times2n$ grid constructed from
                 \begin{equation}
                         \renewcommand{\arraystretch}{1.5}
                     \begin{array}{c|c}
                         \langle\Theta_{00}(\tau_1,k)\Theta_{00}(\tau_2,k)\rangle&\langle\Theta^{\rm S}_{00}(\tau_1,k)\Theta^{\rm S}_{00}(\tau_2,k)\rangle\\\hline
                         \langle\Theta_{00}^{\rm S}(\tau_1,k)\Theta^{\rm S}_{00}(\tau_2,k)\rangle&\langle\Theta^{\rm S}(\tau_1,k)\Theta^{\rm S}(\tau_2,k)\rangle
                     \end{array},
                 \end{equation}
                 where $\langle\Theta^{\rm S}_{00}(\tau_1,k)\Theta^{\rm S}_{00}(\tau_2,k)\rangle$ is the symmetric combination of the cross-correlation between $\Theta_{00}$ and $\Theta^{\rm S}$. After diagonalisation, the first half of the eigenvectors refer to the density and the second to the anisotropic stress. The diagonalisation creates orthogonal eigenvectors which are then used as source terms in the {\ttfamily CAMB}~\cite{Lewis:1999bs} linear Einstein-Boltzmann code. The $C_\ell$ are calculated using each individual eigenvector as a source function $C_\ell^i=u_i(k\tau)/(\sqrt{\tau}(k\tau)^\gamma)$, which can be summed to give the total power spectra
		\vskip-15pt
                 \begin{equation}
                 C_\ell=\sum_{i=1}^n\lambda_iC_\ell^i.
                 \end{equation}
		\vskip-5pt
                 \indent By ordering the $\lambda_i$ from largest to smallest, the required accuracy in the $C_\ell$ can be achieved by including relatively few eigenmodes. This can be seen in the middle row of Fig.~\ref{fig:cls}, where there is only about a 10\% difference between using all 512 eigenmodes of a $512\times512$ grid compared to only using 32 eigenmodes when fixing the value of $G\mu$. Also, it can be seen in the top row of Fig.~\ref{fig:cls} that reducing the grid resolution reduces the amplitude of the $C_\ell$. A grid resolution of $128\times128$ is about 5\% lower, on average, than using the $512\times512$ grid but convergence times decrease drastically. It should be noted that there is negligible difference between using a $512\times512$ and a $1024\times1024$ grid meaning that the former is reliably giving the full $C_\ell$ contribution. The bottom row shows what happens when using more $k$ values in the calculation. Wiggly features arise from using too few $k$ values and can be removed at the expense of a much longer calculation. Using these findings we can choose the optimal UETC parameters to give a good quality $C_\ell$ in a reasonable amount of time. The resulting spectra obtained from our analytical method are shown in Fig.~\ref{fig:cls_cmbact} in green dot-dot-dashed curves and agree well with USM string realisations, especially in the limit of large numbers of simulated segments.\\*    
                 \\*

	\subsection{Comparison of the string power spectrum}
	\label{sec:comp}
\begin{figure}
 \centering
 \includegraphics{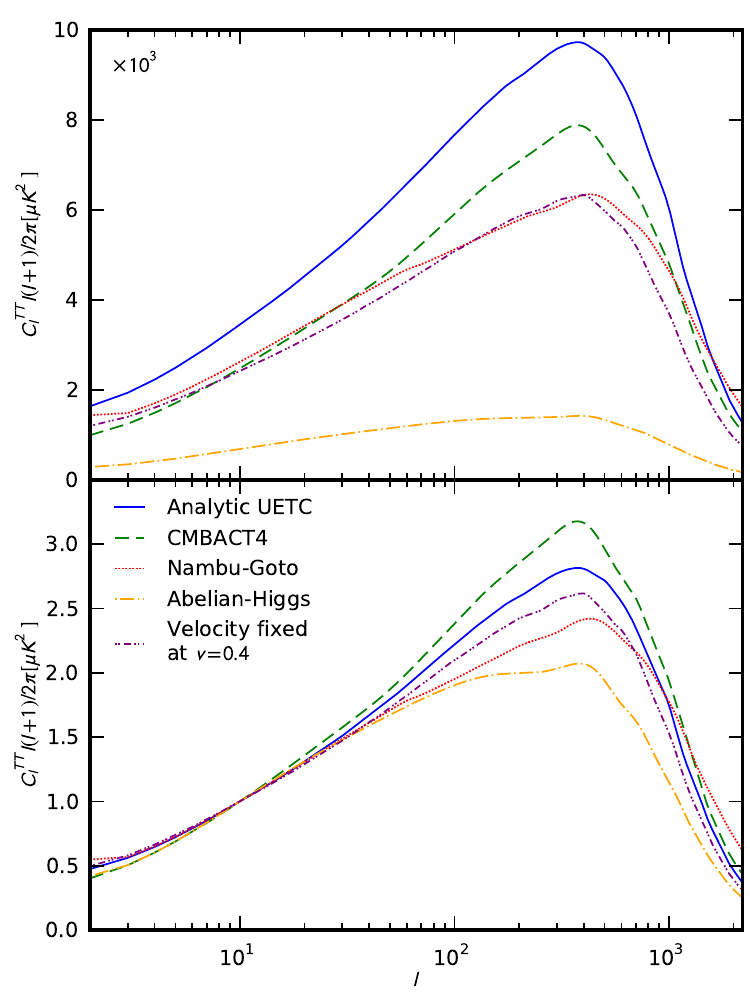}
 \vskip-15pt
 \caption{Comparison of approaches to string modelling, scaled by $G\mu$ in the upper subplot and normalising the temperature power spectrum at $\ell=10$ in the lower subplot. We compare our approach (in solid blue) to {\tt CMBACT4}~\cite{Pogosian:1999np}, Nambu-Goto simulations~\cite{Lazanu:2014xxa}, and Abelian-Higgs simulations~\cite{Bevis:2010gj} (in dashed green, dotted red, dot-dashed orange, and, for the analytic USM with the velocity fixed at $v=0.4$, in dot-dot-dashed purple respectively).}
 \label{fig:methods}
 \vskip-15pt
 \end{figure}

\indent In Fig.~\ref{fig:methods} we compare our temperature power spectrum (scaled by $G\mu$ in the upper subplot and normalised at $\ell=10$) to that of {\tt CMBACT4}~\cite{Pogosian:1999np}, Nambu-Goto simulations~\cite{Lazanu:2014xxa}, and Abelian-Higgs simulations~\cite{Bevis:2010gj}. Both  {\tt CMBACT4} and our method use the same velocity dependent one-scale model parameters, but  {\tt CMBACT4} uses $L_{\rm f}=0.5$. The Nambu-Goto simulations are performed in an expanding background from recombination to today, including $\Lambda$ domination. Large loops are kept in the simulation and contribute to the total energy-momentum tensor of the network, but these simulations cannot resolve small-scale physics near the string width and do not include the effects of radiation backreaction. In contrast, the Abelian-Higgs simulations can resolve small-scale structure and radiative effects~\cite{Daverio:2015nva}. These, however, have a smaller dynamical range and cannot easily evolve through the radiation-matter transition (so the UETC is instead interpolated), but see recent progress in~\cite{Daverio:2015nva} where the authors simulate through the transition.\\* 
\\*
\indent Overall, when normalised at $\ell=10$, the four spectra agree reasonably well. The USM variants ({\tt CMBACT4} and our approach) both predict slightly more power at the peak than either of the simulations. The Nambu-Goto simulations predict more power on very small scales, around twice as much as the Abelian-Higgs model. It is well known that Nambu-Goto calculations yield higher string densities than field theoretic ones, which will increase their overall normalisation. The resulting constraints on $G\mu$ are therefore around a factor of $50\%$ lower~\cite{Ade:2015xua} as can be inferred from the upper subplot in Fig.~\ref{fig:methods}. The USM variants are closer to the Nambu-Goto simulations in this respect~\cite{Lazanu:2014xxa}. In this paper we will not consider using the analytic USM to mimic the Abelian-Higgs spectra. As we have shown, there is some additional uncertainty in the USM, as the normalisation depends somewhat on the choice of $L_{\rm f}$.\\* 
\\*
\indent In summary, given the large differences in modelling between the various approaches we find this comparison encouraging, although more work is needed to further delineate the differences. In particular, as discussed at the end of Sec.~\ref{sec:VOS}, the VOS rms velocity is defined through a worldsheet integral over all scales and receives a large contribution from relativistic wiggles on the string. On the other hand, the USM assumes straight segments moving at a given speed and the small-scale structure on the segments is captured via a ``renormalisation" of their tension. This implies that the speed to be associated to the USM segments must be lower than the VOS rms velocity, and should correspond to the network velocity at correlation length scales. Numerical simulations show this to be significantly lower than the rms speed. This issue has not been examined before, partly because the calculated string spectra from different approaches can differ by up to a factor of 2, and partly because it can be offset by choosing a lower value for the USM parameter $L_{\rm f}$ (see below).  As quantitative agreement between the different approaches is now being established, it is important to fully understand this issue. To this end it will be important to extract the network velocity distribution as a function of length scale in both Nambu-Goto and Abelian-Higgs simulations.\\*  
\\*
\indent Plotted in Fig.~\ref{fig:methods} in purple dot-dot-dash is the $C_\ell$ obtained when $v=0.4$. As can be seen, the peak of the velocity fixed $C_\ell$ has a very similar amplitude to the Nambu-Goto simulation $C_\ell$ in dotted red, although the simulations still have larger power at both lower and higher $\ell$. This supports the idea that the discrepancy in the amplitude of string spectra could be related to different predictions/assumptions on string velocity in the different approaches (cf. the discussion at the end of Sec.~\ref{sec:VOS}). Note that the parameter $L_{\rm f}$ in the USM is somewhat degenerate with the string velocity - for a fixed $v$ a lower $L_{\rm f}$ reduces the density of strings by increasing the string decay rate, thus reducing the $C_\ell$ amplitude and matching simulations better than using $L_{\rm f}=1$. In the absence of a more complete quantitative understanding of the string velocity distribution - input required from string evolution simulations - our string spectra obtained from the USM have a larger amplitude (see the solid blue line in the upper subplot of Fig.~\ref{fig:methods}). This leads to slightly tighter constraints on cosmic strings than in numerical simulations. Marginalising over the network parameters $c_{\rm r}$ and $\alpha$, partly takes care of the differences between $L_{\rm f}=0.5$ and $L_{\rm f}=1$ in the USM since a high $c_{\rm r}$ reduces the velocity [as seen from Eq.~(\ref{VOSv_comoving}) and pictorially in Fig.~\ref{fig:ev}].

   \section{Cosmic Superstrings}
        \label{sec:superstrings}
        
\begin{figure}
 \centering
 \includegraphics{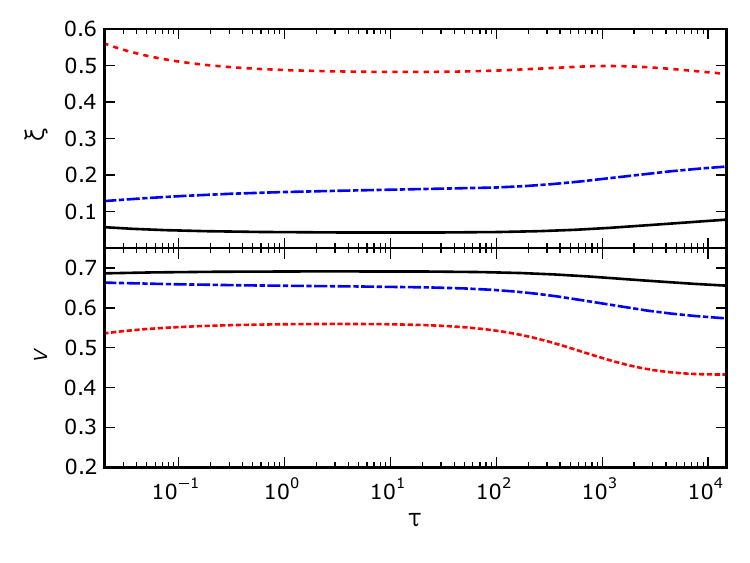}
 \vskip-15pt
 \caption{The radiation and matter era evolution of the velocity $v$, and correlation length $\xi$, for the F-string in solid black, the D-string in dot-dashed blue, and the FD-string in dotted red. These results are obtained when $g_{\rm s}=0.3$, $w=1$, and $c_{\rm s}=0.23$.}
 \label{fig:sev}
\vskip-15pt
 \end{figure}

            A cosmic superstring network can be modelled as a collection of string segments of different types, each string type having its own tension and self-intercommuting probability~\cite{Jones:2003da,Jackson:2004zg,Hanany:2005bc,Jackson:2007hn,Avgoustidis:2005nv,Copeland:2005cy,Tye:2005fn,Hindmarsh:2006qn,Avgoustidis:2007aa,Urrestilla:2007yw,Pourtsidou:2010gu}. Strings of different types interact with each other via ``zipping" or ``unzipping" leading to heavier or lighter strings respectively that are connected to the original strings at trilinear Y-shaped junctions~\cite{Polchinski:2004ia}. The fundamental building blocks for these networks are light (fundamental) F-strings and heavier (Dirichlet) D-strings, with a tension hierarchy controlled by the fundamental string coupling~\cite{Schwarz:1995dk,Witten:1995im,Polchinski:2004ia}. Heavier strings arise as bound states between $p$ F-strings and $q$ D-strings, where $p$,$q$ are coprime. Given the fundamental string tension, the corresponding tensions of these heavier $(p,q)$-strings are controlled mainly by $p$,$q$ and the value of the string coupling. These networks generally behave very differently than their ordinary cosmic string counterparts. They are typically characterised by small intercommutation probabilities, thus leading to higher string number densities~\cite{Jones:2003da,Avgoustidis:2005nv,Tye:2005fn,Pourtsidou:2010gu}. The complex interactions present imply that several string types with different tensions and correlation lengths can simultaneously contribute to the string network CMB spectra.\\*           
         \\*    
		\indent In scaling superstring networks, the string number density is dominated by the lightest F-strings, followed by D-strings and the first bound state, i.e. (1,1)-strings. Heavier bound states are suppressed, so the number of string types considered in the model can be truncated at a finite number. Following~\cite{Pourtsidou:2010gu} we shall describe the network by keeping seven distinct types of strings:  
            
            \begin{align}
            1&\phantom{space}F&(1,0),\nonumber\\
            2&\phantom{space}D&(0,1),\nonumber\\
            3&\phantom{space}FD&(1,1),\nonumber\\
            4&\phantom{space}FFD&(2,1),\nonumber\\
            5&\phantom{space}FDD&(1,2),\nonumber\\
            6&\phantom{space}FFFD&(3,1),\nonumber\\
            7&\phantom{space}FDDD&(1,3),\label{eq:stringtypes}\\
            \vdots&\phantom{spaceFD}\vdots&\vdots\phantom{0),}\nonumber
            \end{align}
            where the last column describes the $(p,q)$ charges of the corresponding string type.\\*
            \\*
	     \indent The large-scale dynamics is then modelled by seven copies of the VOS equations, appropriately extended to account for transfer of energy among the different string types through zipping and unzipping interactions~\cite{Tye:2005fn,Avgoustidis:2007aa}. In each copy of the VOS equations describing a single string, say of type $i$, the self interaction coefficient $c_{\rm r}$ in Eq.~(\ref{VOSxi_comoving}) is replaced by the corresponding self-interaction coefficient $c_i$, and new cross-interaction terms with coefficients $d_{ij}^k$ are added to describe zipping and unzipping. The coefficients $c_i$, $d_{ij}^k$ are controlled by the corresponding microphysical intercommuting probabilities $\mathcal{P}_{ij}$~\cite{Pourtsidou:2010gu}, which can be estimated~\cite{Jackson:2004zg,Jackson:2007hn} from the corresponding string theoretic amplitudes (and field theory approximations in the case of non-perturbative interactions between heavy strings~\cite{Hanany:2005bc}). They can be expressed as a product of two pieces: one that is dependent on the volume of the compact extra dimensions $\mathcal{V}_{ij}(w,g_{\rm s})$, and a quantum interaction piece $\mathcal{F}_{ij}(v,\theta,g_{\rm s})$. Physically, one can think of $\mathcal{V}_{ij}$ as arising from string position fluctuations around the minimum of a localising potential well, giving rise to an effective volume seen by each type of string. The heavier the string the smaller the fluctuations are and thus the smaller the value of $\mathcal{V}_{ij}$~\cite{Jackson:2004zg}. The parameter $w$ corresponds to the effective volume in the compact extra dimensions seen by F-strings. $g_{\rm s}$ is the fundamental string coupling and $v$ and $\theta$ are the relative velocity and angle of the incoming strings. For a pair of strings colliding at an angle $\theta$ and relative speed $v$, the intercommuting probability is
            \begin{equation}
                \mathcal{P}_{ij}(v,\theta,w,g_{\rm s})=\mathcal{F}_{ij}(v,\theta,g_{\rm s})\mathcal{V}_{ij}(w,g_{\rm s}).
            \end{equation}
            Details of how $\mathcal{F}_{ij}$ and $\mathcal{V}_{ij}$ are calculated can be found in~\cite{Pourtsidou:2010gu}. Since the network contains a large number of individual strings with a range of velocities and orientations, the coefficients $c_i$ and $d_{ij}^k$ are determined by the integral of $\mathcal{P}_{ij}$ over a Gaussian velocity distribution centred on the scaling network velocities of each string type and over all angles. This gives the average intercommuting probabilities $\mathcal{P}_{ij}(w,g_{\rm s})\equiv P_{ij}$. Numerical simulations of single-type Nambu-Goto strings with small intercommuting probability~\cite{Avgoustidis:2005nv} suggest that the self-interaction coefficients $c_i$ scale as
            \begin{equation}
                c_i=c_{\rm s}\times P_{ii}^{1/3},
            \end{equation}
            where $c_{\rm s}$ is the standard self-interaction coefficient in three dimensions corresponding to the value $c_{\rm r}$ in Sec.~\ref{sec:VOS}. This choice of $c_{\rm s}$ implies a convenient normalisation of the coefficients $c_i$ so that one recovers the ordinary cosmic string value $c_{\rm r}$ when $P_{ii}=1$. This facilitates direct comparison with ordinary cosmic strings.\\* 
            \\*
            \indent For cross-interactions between two strings of types $i$ and $j$ ($i\ne j$), producing a segment of type $k$, there is an additional factor arising from the kinematic constraints of Y-junction formation~\cite{Copeland:2006if,Copeland:2007nv} that we denote as $S_{ij}^k$ $(i\ne j)$. This also arises as an integral over relative velocities and string orientations~\cite{Avgoustidis:2009ke,Pourtsidou:2010gu}:
            \begin{align}
                S_{ij}^k=&\frac{1}{\mathcal{S}}\int_0^1v^2 dv\int_0^{\pi/2}\sin\theta d\theta \nonumber\\&\phantom{space}\times\Theta(-f_{\overset{\rightarrow}{\mu}}(v,\theta))\exp[(v-\bar{v}_{ij})^2/\sigma_v^2]
            \end{align}
            where $\mathcal{S}$ is a normalisation factor~\cite{Pourtsidou:2010gu}, $\Theta(-f_{\overset{\rightarrow}{\mu}}(v,\theta))$ imposes the kinematic constraints~\cite{Copeland:2007nv}, and $\sigma_v^2$ is the variance of the velocity distribution peaked on the relative scaling velocities $\bar{v}_{ij}=(v_i^2+v_j^2)^{1/2}$ between strings of type $i$ and $j$. The cross-interaction coefficients are then given by 
            \begin{equation}
                d_{ij}^k=d_{ij}\times S^k_{ij}
            \end{equation}
            where $d_{ij}=\kappa\times P_{ij}^{1/3}$. The overall normalisation $\kappa$ is the analogue of $c_{\rm s}$, but for cross-interactions. There is no obvious choice for this phenomenological parameter, but it may be expected to be of order unity by analogy to the ordinary self-interacting string result for $c_{\rm r}$, obtained by numerical simulations. Strictly speaking it should be treated as an extra parameter for the model but, given the large computational resources required in our MCMC analysis, we will set it to unity in this work. Our analysis will still indirectly capture the effects of changing this parameter as it is somewhat degenerate with $w$. To see this, note that $d_{ij}$ is also proportional to $P_{ij}^{1/3}$, which depends weakly on $w$ through the volume factor $\mathcal{V}_{ij}(w,g_{\rm s})$. The leading effect of $w$ is to change the relative amplitude between self-interactions (FF interactions having the strongest $w$ dependence) and cross-interactions of heavy strings, thus mimicking somewhat the effect of varying $\kappa$ relative to $c_{\rm s}$. As computational power improves and our methodology is refined, $\kappa$ should be reintroduced as an additional MCMC parameter.\\* 
            \\*
            \indent The modified VOS equations~\cite{Avgoustidis:2007aa,Pourtsidou:2010gu}, in comoving units, are
            \begin{widetext}
            \vskip-0.6cm
            \begin{align}
            \xi_i'=&\phantom{+}\frac{1}{2\tau}\bigg[2v_i^2\xi_i\tau\frac{a'}{a}-2\xi_i+c_iv_i+\sum_{a,b}\bigg(\frac{d_{ia}^b\bar{v}_{ia}\xi_i\ell^b_{ia}}{\xi_a^2}-\frac{d^i_{ab}\bar{v}_{ab}\xi_i^3\ell^i_{ab}}{2\xi_a^2\xi_b^2}\bigg)\bigg],\label{eq:sxi}\\
            v'=&\phantom{+}\frac{v^2-1}{\tau}\bigg[2v_i\tau\frac{a'}{a}-\frac{k_i}{\xi_i}-\sum_{a,b}b_{ab}^i\frac{\bar{v}_{ab}}{2v_i}\frac{(\mu_a+\mu_b-\mu_i)}{\mu_i}\frac{\xi_i^2\ell^i_{ab}}{\xi_a^2\xi_b^2}\bigg],\label{eq:svel}
            \end{align}
            \end{widetext}
            where $\ell_{ab}^i$ is the average length of segments of type $i$ formed by the zipping/unzipping of string types $a$ and $b$ at conformal time $\tau$, and $\mu_i$ is the tension of the $i^{\rm th}$ string type. All string tensions can be expressed in terms of the fundamental string tension $\mu_{\rm F}$, and in flat spacetime~\cite{Polchinski:2004ia,Schwarz:1995dk,Witten:1995im} are given by: 
            \begin{equation}
                \mu_i=\frac{\mu_{\rm F}}{g_{\rm s}}\sqrt{p_i^2g_{\rm s}^2+q_i^2},
            \end{equation}
 where $p_i$ and $q_i$ are the charges of string type $i$ as listed in (\ref{eq:stringtypes}). The coefficients $b_{ab}^i$ appearing in the velocity evolution equations (\ref{eq:svel}) are related to energy conservation and allow for the energy saved from zipping interactions to be redistributed as kinetic energy of the new segment ($b_{ab}^i=d_{ab}^i$)~\cite{Avgoustidis:2007aa} or radiated away ($b_{ab}^i=0$) as in \cite{Tye:2005fn}. A more realistic model should have a specific radiation mechanism so that $0<b_{ab}^i<d_{ab}^i$, such that some of the energy is redistributed whilst the rest is radiated away. However, for cosmic superstring networks (for which $d_{ij}$ are much smaller than unity) this term has a negligible impact on the string scaling densities and velocities~\cite{Avgoustidis:2009ke,Pourtsidou:2010gu}, so here we take $b_{ab}^i=0$.\\*
 \\*
 \indent Once the velocities and correlation lengths of all string types in the network are obtained by solving (\ref{eq:sxi}) and (\ref{eq:svel}), their unequal-time correlators can be calculated independently as laid out in Sec.~\ref{sec:UETC}. Although $N>3$ string types are needed in order to accurately construct the abundances of the dominant three lighter strings [in this case seven string types are used (\ref{eq:stringtypes})], the resulting scaling densities of the higher charged states with $N>3$ are strongly suppressed compared to the lighter F-, D-, and FD-strings~\cite{Tye:2005fn,Avgoustidis:2007aa,Avgoustidis:2009ke}. This allows us to only consider these first three states in the computation of CMB signatures through our UETC analytic method. The evolution of the network parameters for the three lightest strings can be seen in Fig.~\ref{fig:sev} for $c_{\rm s}=0.23$, $w=1$, and $g_{\rm s}=0.3$. \\*
    \\*
    \indent Once the UETC of each of the three lighter strings are calculated they can simply be summed to give the total string UETC, since the individual segments are uncorrelated in the USM. This can then be diagonalised and the eigenvectors and eigenmodes used as sources for finding the contribution from cosmic superstrings to the CMB anisotropy. We have checked that our analytic UETC method reproduces the results of Fig.~4 in~\cite{Pourtsidou:2010gu}, including the shift in the location of the peak as we vary $g_{\rm s}$. We have found a slightly lower amplitude in the B-mode spectrum that can be attributed to the extra factor of 2 in the vector modes that was present in {\ttfamily CMBACT3} (which~\cite{Pourtsidou:2010gu} was based on) and has been corrected in {\ttfamily CMBACT4}~\cite{Pogosian:2014}.
     
   \section{String Constraints}
        \label{sec:constraints}
        
\begin{figure*}
                     \centering
                     \includegraphics{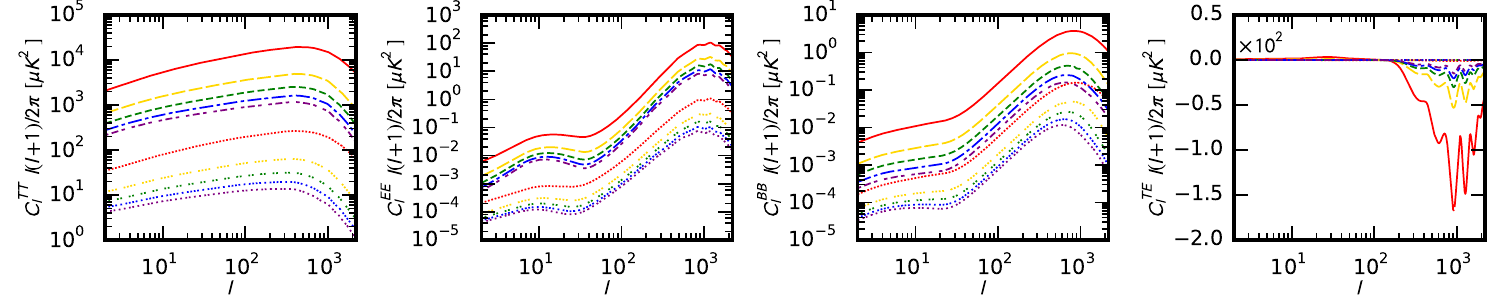}
                     \vskip-10pt
                     \caption{The total $C_\ell$ (scalar+vector+tensor modes) for different values of $c_{\rm r}$ and $\alpha$. The red solid lines show $c_{\rm r}=0.1$ and through yellow (long-dashed), green (short-dashed), blue (dot-dashed), and purple (dot-dash-dotted) for $c_{\rm r}=0.3$, $c_{\rm r}=0.5$, $c_{\rm r}=0.7$, and $c_{\rm r}=0.9$. The upper (solid-patterned) lines indicate $\alpha=10$ whilst the lower (dotted versions of the pattern) lines are for $\alpha=1$.  This is shown for $C_\ell^{TT}$, $C_\ell^{EE}$, $C_\ell^{BB}$, and $C_\ell^{TE}$ in columns 1-4.}
                     \label{fig:cralpha}
                     \vskip-5pt
                  \end{figure*}

\begin{figure*}
                     \centering
                     \includegraphics{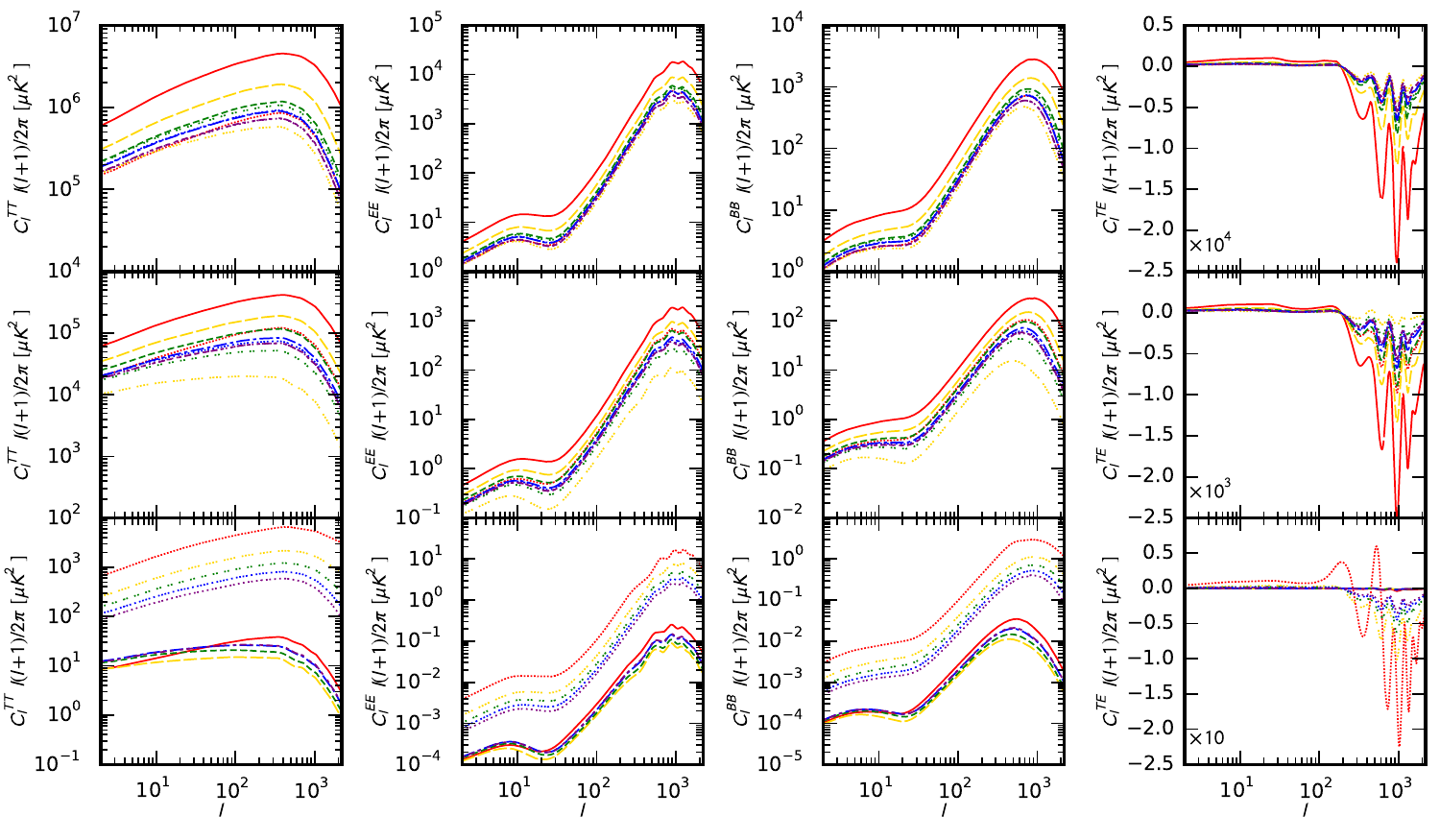}
                     \vskip-10pt
                     \caption{The total $C_\ell$ obtained from cosmic superstrings for different values of $g_{\rm s}$, $c_{\rm s}$ and $w$. As with previous figures, the first column shows the temperature auto-correlation $C_\ell$, the second the EE, the third the BB and the last column shows the temperature, E-mode correlation $C_\ell$. The three rows show $g_{\rm s}$ values of $10^{-2}$, $10^{-1}$, and $0.9$ from top to bottom. The colouring and patterning system is the same as in Fig.~\ref{fig:cralpha} with red (solid), yellow (long-dashed), green (short-dashed), blue (dot-dashed) and purple (dot-dash-dotted) lines indicating the values of $c_{\rm s}$ from $c_{\rm s}=0.1$ to $0.9$ in steps of $0.2$. The width of the band of similar colour and pattern indicates the upper and lower $w$ values with the dotted-patterned lines defined by $w=10^{-3}$ and the solid-patterned lines by $w=1$.}
                     \label{fig:cralpha1}
                     \vskip-5pt
                  \end{figure*}

\begin{figure*}
            \centering
            \includegraphics{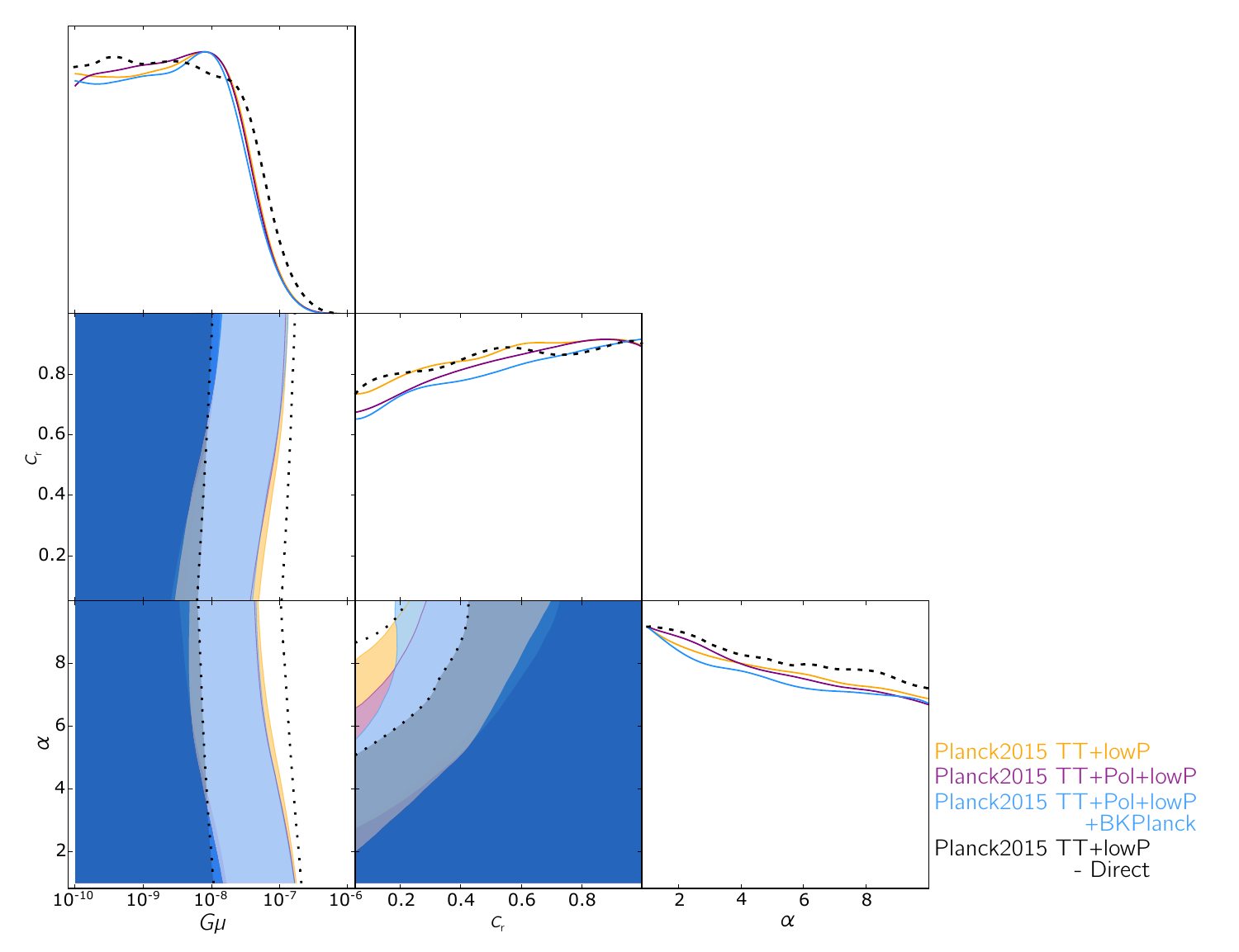}
            \caption{$2\sigma$ likelihood contours for $G\mu$, $c_{\rm r}$ and $\alpha$ from the string $C_\ell$ interpolation and direct calculation methods. The orange line shows the constraints from Planck2015 TT+lowP, purple and blue lines are used for Planck2015 TT+Pol+lowP and Planck2015 TT+Pol+lowP+BKPlanck respectively. The black dashed line shows the direct calculation constraints for Planck2015 TT+lowP.}
            \label{fig:ssconstraints}
        \end{figure*}

\begin{figure*}
            \centering
            \includegraphics{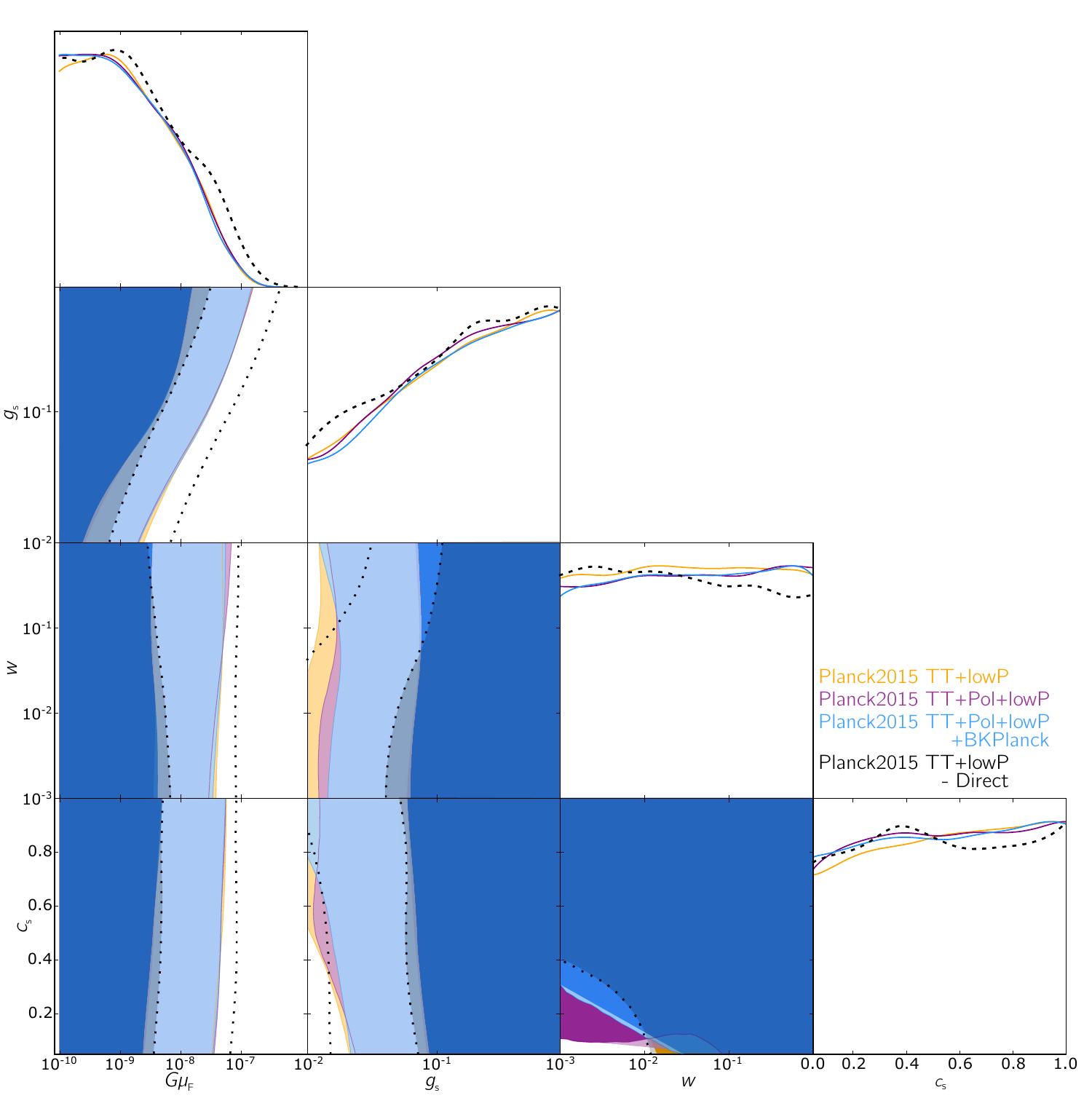}
            \caption{$2\sigma$ likelihood obtained for $G\mu_{\rm F}$, $g_{\rm s}$, $w$ and $c_{\rm s}$. The black dashed line (and black shading) shows the constraints from the Planck2015 TT+lowP direct calculation method and orange, purple, and blue lines are the Planck2015 TT+lowP, Planck2015 TT+Pol+lowP and Planck2015 TT+Pol+lowP+BKPlanck constraints using the interpolation method.}
            \label{fig:sssconstraints}
        \end{figure*}

        We obtain joint constraints on cosmic string network and $\Lambda$CDM parameters using a modified version of {\ttfamily COSMOMC}. To reduce computational time in our analysis we have tested two methods for deriving string network constraints. In the first method, the string $C_\ell$ are pre-calculated for ranges $c_{\rm r}=[0.1,1]$ and $\alpha=[1,10]$ at the Planck best fit values for the cosmological parameters, i.e. $\Omega_{\rm b}h^2$, $\Omega_{\rm c}h^2$, and $H_0$. These $C_\ell$ are read into {\ttfamily COSMOMC}, interpolated at the MCMC $c_{\rm r}$ and $\alpha$ values and then scaled by $(G\mu)^2$. This is an extremely efficient way for obtaining network constraints since only the $\Lambda$CDM $C_\ell$ need to be calculated, while the interpolation takes very little time. We have checked that the difference in the resulting string $C_\ell$ when calculated at the upper and lower $3\sigma$ bounds in $\Omega_{\rm b}h^2$, $\Omega_{\rm c}h^2$, and $H_0$ is $\sim\!0.5\%$ in the temperature, E- and B-modes and no more than $\!\sim10\%$ in the TE cross-correlation. This uncertainty in the string $C_\ell$ is $\ll1\%$ of the total $C_\ell$. The $C_\ell$ for different $c_{\rm r}$ and $\alpha$ are plotted in Fig.~\ref{fig:cralpha}. The different bands of colour indicate the value of $c_{\rm r}$, solid red being the lowest ($c_{\rm r}=0.1$), then progressing through long-dashed yellow, short-dashed green, dot-dashed blue and dot-dash-dotted purple in steps of 0.2, up to $c_{\rm r}=0.9$. The upper (patterned) and lower (dot-patterned) edges of the bands indicate $\alpha=10$ and $\alpha=1$ respectively. From this it can be seen that the effect of $\alpha$ is to change the amplitude of the $C_\ell$, with a lower $\alpha$ also flattening the small $\ell$ features (as best seen in the second column and to a lesser extent in the third column of Fig.~\ref{fig:cralpha}). Increasing $c_{\rm r}$ reduces the amplitude of the $C_\ell$ and, as best seen in the third column of Fig.~\ref{fig:cralpha}, shifts the main peak towards slightly smaller $\ell$. In the second method, which is computationally expensive, we simply calculate the string and $\Lambda$CDM $C_\ell$ for each (network) parameter value and compare it to the CMB data. \\*  
                    \\*
                  \indent The same process of pre-calculating string spectra can be done for cosmic superstring networks in the parameter ranges $c_{\rm s}=[0.1,1]$, $g_{\rm s}=[0.01,0.9]$ and $w=[0.001,1]$. The superstring $C_\ell$ can be seen in Fig.~\ref{fig:cralpha1}, where same colours and patterns are used for the steps in $c_{\rm s}$ as in Fig.~\ref{fig:cralpha}. The bands indicate values of $w$, with $w=10^{-3}$ corresponding to the solid-patterned lines and $w=1$ to the dotted version of the same pattern. The rows indicate varying values of $g_{\rm s}$, with $g_{\rm s}=0.01$, $g_{\rm s}=0.1$, and $g_{\rm s}=0.9$ for the top, middle and bottom rows respectively. The first point to notice is that the $C_\ell$ amplitudes at low $g_{\rm s}$ are much greater than those at large $g_{\rm s}$. For large $c_{\rm s}$ values there is less difference between the greatest and smallest values of $w$, especially at low $g_{\rm s}$, i.e. the purple dot-dash-dotted lines in the top row of Fig.~\ref{fig:cralpha1} overlap, but are well separated in the bottom row. This is because for large $c_{\rm s}$ the cross-interaction terms $d_{ij}^k$ (which are less dependent on $w$ than the self-interaction terms $c_{i}$) play a more important role in setting the scaling string number densities. For small values of $c_{\rm s}$, the $c_{i}$ coefficients become smaller (while $d_{ij}^k$ are unaffected) leading to small correlation lengths and so large string number densities. The $C_\ell$ amplitudes are then affected more strongly by $c_{i}$, giving rise to a stronger dependence on $w$. 
\\*
\\*
        \indent The datasets used in the MCMC analysis come from the Planck2015 mission~\cite{Aghanim:2015xee}, in particular: 
        {\bf Planck2015 TT+lowP}: This contains the 100-GHz, 143-GHz, and 217-GHz binned half-mission TT cross-spectra for $\ell=30-2508$ with CMB-cleaned 353-GHz map, CO emission maps, and Planck catalogues for the masks and 545-GHz maps for the dust residual contamination template. It also uses the joint temperature and the E and B cross-spectra for $\ell=2-29$ with E and B maps from the 70-GHz LFI full mission data and foreground contamination determined by 30-GHz Low Frequency Instrument (LFI) and 353-GHz High Frequency Instrument maps. \\\
        {\bf Planck2015 TT+Pol+lowP}: This contains the same data as Planck2015 TT+lowP but also uses the TE and EE cross-spectra for $\ell=30-1996$.\\
        {\bf Planck2015 TT+Pol+lowP+BKPlanck}: This again contains all of the data used in Planck2015 TT+Pol+lowP but includes also the cross-frequency spectra between BICEP2/Keck maps at 150 GHz and Planck maps at 353 GHz including the B-mode spectra at multipoles $\ell\sim50-250$.\\*
        \\*
        \indent We first consider our interpolation method, where the $C_\ell$ are pre-calculated on a grid in $c_{\rm r}$ and $\alpha$ (or in the case of cosmic superstring networks $c_{\rm s}$, $g_{\rm s}$, and $w$), and then a spline interpolation used between grid values. The results obtained from this method are very quick and accurate due to our ability to use all 512 eigenmodes of the $512\times512$ grid for the UETC. The constraints on network parameters derived from this method are shown in Fig.~\ref{fig:ssconstraints}. $G\mu$ is implemented into the MCMC analysis through a logarithmic prior of $[-10,-5]$ such that $G\mu=10^{[-10,-5]}$.\\*
        \\*
        \indent There is no significant difference in our constraints when using Planck2015 TT+lowP, or including EE and TE or both EE and TE and BB results. The upper $2\sigma$ value for the tension is $G\mu<1.1\times10^{-7}$ for Planck2015 TT and is similarly $G\mu<9.6\times10^{-8}$ and $G\mu<8.9\times10^{-8}$ for Planck2015 TT+Pol+lowP and Planck2015 TT+Pol+lowP+BKPlanck. These agree well with the $G\mu<1.8\times10^{-7}$ and $G\mu<1.3\times10^{-7}$ from the Planck cosmological parameters analysis~\cite{Ade:2015xua}. The slightly tighter constraints obtained here are due to the amplitude of the $C_\ell$ not scaling with the value of $L_{\rm f}$, i.e. the $C_\ell$ are larger when $L_{\rm f}=1$ as assumed here, while previous results were obtained from {\tt CMBACT} with  $L_{\rm f}=0.5$. There is little difference between using the Planck temperature data alone and including polarisation data as expected from~\cite{Ade:2015xua}.  As can be seen in the other two columns of Fig.~\ref{fig:ssconstraints}, $c_{\rm r}$ and $\alpha$ are not constrained. There is a slight preference for higher values of $c_{\rm r}$ and lower values of $\alpha$ since both of these lead to a smaller $C_\ell$. Features, such as the position of the main peak or the pronounced lower $\ell$ peak make very little difference to the overall constraints. There is a very slight correlation between $G\mu$ and $c_{\rm r}$ and anti-correlation between $G\mu$ and $\alpha$, as expected from the $C_\ell$ seen in Fig.~\ref{fig:cralpha}. A combination of high $\alpha$ and low $c_{\rm r}$ is mildly disfavoured. Further, by comparing the constraints on $G\mu$ and $c_{\rm r}$ to their effect on the $C_\ell$ in Fig.~\ref{fig:cralpha} there is a larger difference between changes at small $c_{\rm r}$ than changes at large $c_{\rm r}$. For this reason we expect to see a greater correlation between $G\mu$ and $c_{\rm r}$ on a logarithmic scale from values $c_{\rm r}\ll1$ to $c_{\rm r}\approx0.1$ than implied over our prior range.\\*
        \\*
        \indent Considering our direct calculation method, where the string spectra are calculated every time along with the $C_\ell$ from $\Lambda$CDM, the constraints are slightly weaker. This is because there is a pay-off between the resolution of the UETC and the number of eigenmodes used in the reconstruction and the time spent computing the spectra. To efficiently calculate the constraints a grid resolution of $128\times128$ with 64 eigenmodes has been used. As can be seen in Fig.~\ref{fig:cls} we expect a reduction in power of about $10-20\%$ which means the value of $G\mu$ is allowed to be higher than when the high resolution, full reconstruction interpolation method is used. For Planck2015 TT+lowP this is $G\mu<4.3\times10^{-7}$. The constraints on $c_{\rm r}$ and $\alpha$ also show a slight preference for lower $c_{\rm r}$ and larger $\alpha$, as in our interpolation method.\\*
        \\*
        \indent For cosmic superstrings, $G\mu_{\rm F}$, $g_{\rm s}$ and $w$ are marginalised over logarithmic priors, and $c_{\rm s}$ over a flat prior. Again all 512 eigenmodes of the $512\times512$ grid for the UETC are used. The likelihood contours obtained from our interpolation method can be found in Fig.~\ref{fig:sssconstraints}. It can be seen that $w$ and $c_{\rm s}$ are almost flat (columns 3 and 4), again with larger values of $c_{\rm s}$ favoured as this leads to a smaller amplitude $C_\ell$. As the string density grows with decreasing $g_{\rm s}$, the constraints on $g_{\rm s}$ favour larger values, as seen in the second column. Note, however, that the model is not reliable for large values of $g_{\rm s}$ as the perturbative expansion starts to break down and the string interaction amplitudes used in $c_i$ and $d_{ij}^k$ have large uncertainties. Finally, the first column shows our constraints on the fundamental string tension $G\mu_{\rm F}$, which is much smaller than for ordinary cosmic strings. We find $G\mu_{\rm F}<2.8\times10^{-8}$ for Planck2015 TT+lowP when marginalising over $g_{\rm s}$, $c_{\rm s}$, and $w$, and the same constraint for Planck2015 TT+Pol+lowP and Planck2015 TT+Pol+lowP+BKPlanck.\\* 
        \\*
        \indent Also in Fig.~\ref{fig:sssconstraints} we show the constraints when using the direct calculation method, where the string spectra are calculated at every step in the Markov chain. This is a much more intensive computation and so a lower resolution grid and fewer eigenmodes in the reconstruction had to be used. As for cosmic strings, the optimal balance between computing time and accuracy suggested using a $128\times128$ grid with 64 eigenmodes. The constraints are thus slightly weaker, with the main result $G\mu_{\rm F}<4.2\times10^{-8}$. The results from our two methods are in good agreement, justifying the use of our interpolation method, and showing that varying $\Lambda$CDM parameters within Planck priors has little effect on the string constraints. 
        
 \section{Conclusions}
        \label{sec:discussion}
  
 Currently, there are two main approaches to  the detection of cosmic strings. First, since they actively generate scalar, vector and tensor perturbations they lead to signatures in the temperature, polarisation, and non-Gaussian spectra of the CMB.  Second, a cosmic string network will emit gravitational waves, primarily from loop decay. This leads to a stochastic background which can be constrained using pulsar timing, laser interferometry experiments such as LIGO and eLISA, and also  the CMB~\cite{Smith:2006nka}. A transient gravitational wave signal is also expected from cusps and kinks in the network~\cite{Damour:2004kw}. The latter class of tests has the potential to provide even stronger constraints on the string tension $G\mu$, but there are large uncertainties in the loop size, which is fixed by gravitational back-reaction. Model dependence on gravitational waves from cosmic strings further makes it difficult to determine signatures, for example, whilst Nambu-Goto strings decay into loops, Abelian-Higgs strings primarily decay into particles~\cite{Damour:2004kw,Dufaux:2010cf,Olmez:2010bi}. It is therefore important to use a variety of complementary observational probes.\\* 
  \\*
\indent The first class of tests also suffer from uncertainties, but these are less significant.  The string UETC can be obtained from simulations and used as source functions in CMB codes, but simulations are numerically expensive and suffer from issues in dynamical range. An alternative approach is to model the string network as an ensemble of  segments using the USM. Crucially, although the USM provides a simplified picture of the network, it is able to match simulations by adjusting the free parameters of the model, namely the correlation length, the rms velocity and string wiggliness.\\*
\\*
\indent In this paper we have significantly improved and extended our previous work on string power spectra from the USM:
\begin{enumerate}
\item We have analytically solved the UETC for an evolving string network, whereas our previous work was restricted to constant network parameters. The UETC itself can be computed in under a minute. For the CMB power spectrum, although the time taken is increased due to tracking a larger number of Fourier modes, on a 3.1 GHz Intel Xeon
CPU with eight threads, our code runs in $\sim 60$ minutes. For comparison, around 2000 network realisations are required for {\tt CMBACT4} to achieve the same accuracy and since this code is serial, the computation time is $\sim 30$ hours.
\item We have extended the formalism to cosmic superstring networks with multiple string types and different network parameters.  Here the UETC can be computed for each string type and added, since the segments are assumed to be uncorrelated. The UETC calculation is much quicker than the CMB line-of-sight integration, so the total computation time is not significantly increased over the single string case.
\item For the first time we have been able to marginalise over the string network parameters when fitting to Planck2015 and joint Planck-BICEP2 data. The data is consistent with no strings for either the single or multi-string case. Since other network parameters are unconstrained when the tension is very small, it is only possible to present joint constraints on these with $G\mu$. In the  superstring case, for example, the constraint on the string coupling $g_{\rm s}$ is degenerate with $G\mu_{\rm F}$.
\end{enumerate}

There are several possibilities to explore in future work. First, there are various ways in which the USM could be improved. Superstring networks contain Y-type junctions, but in the present formulation these only impact the evolution of the network parameters. Since junctions are relatively rare in the limit of large and small coupling, the USM is expected to provide a sufficient description. However, in some regimes the energy density of the network may not be dominated by a single string type, and junctions may become important. In this case the USM could be modified to include a correlation between segments. 
 A further improvement is the inclusion of loops. The decay of string segments in the USM should mimic the energy loss in loops, but it is possible these may lead to additional interesting signatures.\\*
  \\*
\indent Given that Planck has largely exhausted the available signal in the temperature data, future string constraints from the CMB will be driven by polarisation and non-Gaussianity. The non-Gaussian signal from post-recombination simulations has been used to obtain constraints on $G\mu$~\cite{Ade:2013xla}, and attempts have been made to compute the bi-spectrum analytically using a Gaussian model for the string correlators~\cite{Regan:2015cfa}. It is also possible to compute the non-Gaussian signal using the USM which will, by design, include physics from recombination and along the line-of-sight. This has already been demonstrated for the CMB bi-spectrum~\cite{Gangui:2001fr} by performing many realisations of the network.  It is possible to employ a similar analytic method used in this work to compute the string bi- and tri-spectrum, which we would expect to be significantly faster~\cite{Charnock:-)}. \\*
 \\*
 \indent The detection of gravitational waves by LIGO is particularly exciting for strings, and the next generation of ground and space based experiments can potentially provide much stronger limits than those from the CMB.  However, these limits strongly depend on modelling, for example, the loop, kink and cusp distribution. Further work is needed to understand these and until then the CMB will continue to be an important tool in the search for strings. 
 
        \section*{Acknowledgements}
        We appreciate our useful conversations with Richard Battye, Levon Pogosian, Carlos Martins, and Jon Urrestilla. We also thank Martin Kunz, Andrei Lazanu and Paul Shellard for the use of their string spectra. T. C. is supported by n STFC studentship. The work of A. A. is supported by an Advanced Research Fellowship at the University of Nottingham. A. M. is supported by a Royal Society University Research Fellowship.  E. J. C. is supported by STFC Consolidated Grant No. ST/L000393/1. We are grateful for our access to the University of Nottingham High Performance Computing Facility. Finally, we thank the referee for her or his invaluable comments on this paper.
    
    \bibliography{stringsbib}
        \begin{widetext}
                \begin{table}[h!]
                \begin{tabular}{rlcrl}
                    $I_1(x,\varrho)=$&$\dfrac{1}{2}\dint_0^\pi d\theta\sin\theta\sec^2\theta\cos(x\cos\theta)J_0(\varrho\sin\theta)$&~~~~~~~~~~~&$I_4(x,\varrho)=$&$\dfrac{1}{2}\dint_0^\pi d\theta\sin\theta\sec^2\theta\cos(x\cos\theta)\frac{J_1(\varrho\sin\theta)}{\varrho\sin\theta}$\\
                   $=$&$\dsum_{c=0}^\infty\dfrac{1}{c!}\dfrac{\varrho}{(2c-1)}\bigg(-\dfrac{x^2}{2\varrho}\bigg)^2j_{c-1}(\varrho)$&~~~~~~~~~~~&$=$&$\dfrac{\cos x}{\varrho^2}-\dsum_{c=0}^\infty\dfrac{1}{c!}\dfrac{1}{(2c-1)}\bigg(-\dfrac{x^2}{2\varrho}\bigg)^2j_{c-2}(\varrho)$\\\\
                    $I_2(x,\varrho)=$&$\dfrac{1}{2}\dint_0^\pi d\theta\sin\theta\cos(x\cos\theta)J_0(\varrho\sin\theta)$&~~~~~~~~~~~&$I_5(x,\varrho)=$&$\dfrac{1}{2}\dint_0^\pi d\theta\sin\theta\cos(x\cos\theta)\dfrac{J_1(\varrho\sin\theta)}{\varrho\sin\theta}$\\
                    $=$&$\bigg(\dfrac{\sin\sqrt{\varrho^2+x^2}}{\sqrt{\varrho^2+x^2}}\bigg)$&~~~~~~~~~~~&$=$&$\dfrac{1}{\varrho^2}\left(\cos x-\cos\sqrt{\varrho^2+x^2}\right)$\\\\
                    $I_3(x,\varrho)=$&$\dfrac{1}{2}\dint_0^\pi d\theta\sin^3\theta\cos(x\cos\theta)J_0(\varrho\sin\theta)$&~~~~~~~~~~~&$I_6(x,\varrho)=$&$\dfrac{1}{2}\dint_0^\pi d\theta\sin^3\theta\cos(x\cos\theta)\dfrac{J_1(\varrho\sin\theta)}{\varrho\sin\theta}$\\
                    $=$&$\bigg[1+\dfrac{\partial^2}{\partial x^2}\bigg]\bigg(\dfrac{\sin\sqrt{\varrho^2+x^2}}{\sqrt{\varrho^2+x^2}}\bigg)$&~~~~~~~~~~~&$=$&$-\dfrac{1}{\varrho^2+x^2}\bigg[1+\dfrac{1}{x}\dfrac{\partial}{\partial x}\bigg]\left(\cos\sqrt{\varrho^2+x^2}\right)$
                \end{tabular}
                \caption{Integral identities for the UETC.}
                \label{tab:integrals}
                \end{table}
                \end{widetext}
\clearpage
\newgeometry{margin=2cm}
        \begin{table}
        \renewcommand{\arraystretch}{2.1}
        \begin{center}
            \begin{tabular}{l||lrlrlrlrl}
                     &$\langle\Theta_{00}(\tau_1,k)\Theta_{00}(\tau_2,k)\rangle$&&$\langle\Theta^{\rm S}(\tau_1,k)\Theta^{\rm S}(\tau_2,k)\rangle$&&$\langle\Theta^{\rm V}(\tau_1,k)\Theta^{\rm V}(\tau_2,k)\rangle$&&$\langle\Theta^{\rm T}(\tau_1,k)\Theta^{\rm T}(\tau_2,k)\rangle$&&$\langle\Theta_{00}^{\rm S}(\tau_1,k)\Theta^{\rm S}_{00}(\tau_2,k)\rangle$\\\hline\hline
                $a_1$    & $2\alpha^2$    && $\dfrac{1}{2\alpha^2}$                                           && 0     && $\dfrac{1}{4\alpha^2}$       && $1$ \\
                $b_1$    & 0              && $1-\dfrac{1}{2\alpha^2}$                                               && 0     && $-\dfrac{1}{4\alpha^2}$   && $-\dfrac{1}{2}+\alpha^2$\\\rule[-1em]{0pt}{0pt}
                $c_1$    & 0              && $\dfrac{1}{2\alpha^2}-2+2\alpha^2-\dfrac{27\alpha^2}{2\varrho^2}$        && $\dfrac{3\alpha^2}{\varrho^2}$            && $\dfrac{1}{4\alpha^2}-\dfrac{3\alpha^2}{4\varrho^2}$   && 0\\\hline
                $a_2$    & 0              && $\dfrac{3}{2\alpha^2}$                                           && 0     && $-\dfrac{1}{4\alpha^2}$   && $-3$    \\
                $b_2$    & 0              && $-\dfrac{3}{2\alpha^2}$                                                           && 0     && $\dfrac{1}{4\alpha^2}$   && $\dfrac{3}{2}-\dfrac{3\alpha^2}{2}$\\\rule[-1em]{0pt}{0pt}
                $c_2$    & 0              && $\dfrac{3}{2\alpha^2}-\dfrac{3\alpha^2}{2}+\dfrac{27\alpha^2}{2\varrho^2}$         && $-\dfrac{3\alpha^2}{\varrho^2}$&& $-\dfrac{1}{4\alpha^2}+\dfrac{3\alpha^2}{4\varrho^2}+\dfrac{\alpha^2}{4}$   &&0    \\\hline
                $a_3$    & 0              && $-\dfrac{9}{2\alpha^2}$                                          && $\dfrac{1}{\alpha^2}$                              &&  $-\dfrac{1}{4\alpha^2}$   && 0\\
                $b_3$    & 0              && $\dfrac{9}{2\alpha^2}-\dfrac{9}{2}$                                                && $1-\dfrac{1}{\alpha^2}$                        && $\dfrac{1}{4\alpha^2}-\dfrac{1}{4}$   && 0\\\rule[-1em]{0pt}{0pt}
                $c_3$    & 0              && $-\dfrac{9}{2\alpha^2}+9-\dfrac{9\alpha^2}{2}$                                     && $\dfrac{1}{\alpha^2}-2+\alpha^2$              && $-\dfrac{1}{4\alpha^2}+\dfrac{1}{2}-\dfrac{\alpha^2}{4}$   && 0\\\hline
                $a_4$    & 0              && 0                                                                && 0     && 0   && 0 \\
                $b_4$    & 0              && $-\dfrac{3}{2}$                                                  && 0     && $\dfrac{1}{4}$   && $-\dfrac{3\alpha^2}{2}$\\\rule[-1em]{0pt}{0pt}
                $c_4$    & 0              && $3-6\alpha^2+\dfrac{27\alpha^2}{\varrho^2}$                      && $\alpha^2-\dfrac{6\alpha^2}{\varrho^2}$    && $-\dfrac{1}{2}+\dfrac{3\alpha^2}{2\varrho^2}$    && 0 \\\hline
                $a_5$    & 0              && 0                                                                && 0     && 0   && 0 \\
                $b_5$    & 0              && $\dfrac{3}{2}$                                                           && 0     && $-\dfrac{1}{4}$   && $\dfrac{3\alpha^2}{2}$ \\\rule[-1em]{0pt}{0pt}
                $c_5$    & 0              && $-3+6\alpha^2-\dfrac{27\alpha^2}{\varrho^2}$                                                           && $-\alpha^2+\dfrac{6\alpha^2}{\varrho^2}$&& $\dfrac{1}{2}-\dfrac{3\alpha^2}{2\varrho^2}$   && 0 \\\hline
                $a_6$    & 0              && 0                                                                && 0     && 0   && 0 \\
                $b_6$    & 0              && $\dfrac{9}{2}$                                                   && $-1$  && $\dfrac{1}{4}$   && 0 \\
                $c_6$    & 0              && $-9+9\alpha^2$                                                   && $2-2\alpha^2$ && $-\dfrac{1}{2}+\dfrac{\alpha^2}{2}$   && 0\\\hline\rule[-1em]{0pt}{0pt}
                $B$      & $\alpha^2x_1x_2$ && $\dfrac{x_1x_2}{5\alpha^2}X$ && $\dfrac{x_1x_2}{15\alpha^2}X$ && $\dfrac{x_1x_2}{15\alpha^2}X$ && 0 \\\hline
                $C$     & $Z^a$ && $YZ^a+\alpha^2v(\tau)^4Z^b$ && $YZ^a+\alpha^2v(\tau)^4Z^b$ &&$YZ^a+\alpha^2v(\tau)^4Z^b$&&$[Y-v(\tau)^4(1-\alpha^2)]Z^a$ \\\hline
                $z_1^a$ & $-2\alpha^2$ && $-\dfrac{2}{\alpha^2}$ && $\dfrac{2}{3\alpha^2}$ && $\dfrac{-2}{3\alpha^2}$ && $-4$ \\
                $z_2^a$ & $2\alpha^2$  && $\dfrac{1}{2\alpha^2}-\dfrac{9}{\alpha^2x^2}$ && $\dfrac{2}{\alpha^2x^2}$ && $\dfrac{1}{4\alpha^2}-\dfrac{1}{2\alpha^2x^2}$ && $1$ \\
                $z_3^a$ & 0            && $-\dfrac{3}{2\alpha^2x}+\dfrac{9}{\alpha^2x^3}$ && $-\dfrac{2}{\alpha^2x^3}$ && $\dfrac{1}{4\alpha^2x}+\dfrac{1}{2\alpha^2x^2}$ && $\dfrac{3}{x}$ \\
                $z_4^a$ & $2\alpha^2$          && $\dfrac{1}{2}$ && 0 && $\dfrac{1}{4\alpha^2}$ && $1$\\
                $z_1^b$ & 0            && $-2$ && 0 && 0 && 0\\
                $z_2^b$ & 0            && $\dfrac{11}{16}-\dfrac{27}{8x^2}$ && $\dfrac{1}{8}+\dfrac{3}{4x^2}$ && $\dfrac{3}{32}-\dfrac{3}{16x^2}$ && 0\\
                $z_3^b$ & 0            && $\dfrac{3}{16x}+\dfrac{27}{8x^3}$ && $\dfrac{1}{8x}-\dfrac{3}{4x^3}$ && $-\dfrac{5}{32x}+\dfrac{3}{16x^3}$ && 0\\
                $z_4^b$ & 0            && $\dfrac{11}{16}$ && $\dfrac{1}{8}$ && $\dfrac{3}{32}$ && 0\\               
            \end{tabular}
            \caption{Coefficients for the amplitude equations given by $A_i = a_i+b_i(v(\tau_1)^2+v(\tau_2)^2)+c_iv(\tau_1)^2v(\tau_2)^2$. The small $x$ approximation and the ETC are expressed in terms of the functions $X=\bigg[1-(v(\tau_1)^2+v(\tau_2)^2)\bigg(1-\dfrac{\alpha^2}{2}\bigg)+v(\tau_1)^2v(\tau_2)^2(1-\alpha^2+\alpha^4)\bigg]$, $Y=\bigg[1-v(\tau)^2(2-\alpha^2)+v(\tau)^4(1-\alpha^2)\bigg]$ and $Z^j=z^j_1+z^j_2\cos x+z^j_3\sin x+z^j_4x{\rm Si}[x]$.}
            \label{tab:amplitudes}
            \end{center}
        \end{table}

\end{document}